\documentclass[amsmath,showpacs,nofootinbib,superscriptaddress]{revtex4-2}
\usepackage{graphicx}
\usepackage{dcolumn}
\usepackage{bm}
\usepackage{subcaption}
\usepackage{color} 
\usepackage{slashed}
\usepackage{float}
\usepackage{amsfonts}
\usepackage{multirow}
\begin{document}
\newcommand{\overlr}{\stackrel{\leftrightarrow}{\partial}}
\newcommand{\hs}{\hspace*{0.5cm}}
\newcommand{\vs}{\vspace*{0.5cm}}
\newcommand{\be}{\begin{equation}}
\newcommand{\ee}{\end{equation}}
\newcommand{\bea}{\begin{eqnarray}}
\newcommand{\eea}{\end{eqnarray}}
\newcommand{\ben}{\begin{enumerate}}
\newcommand{\een}{\end{enumerate}}
\newcommand{\bde}{\begin{widetext}}
\newcommand{\ede}{\end{widetext}}
\newcommand{\nn}{\nonumber}
\newcommand{\crn}{\nonumber \\}
\newcommand{\Tr}{\mathrm{Tr}}
\newcommand{\non}{\nonumber}
\newcommand{\noi}{\noindent}
\newcommand{\al}{\alpha}
\newcommand{\la}{\lambda}
\newcommand{\bet}{\beta}
\newcommand{\ga}{\gamma}
\newcommand{\va}{\varphi}
\newcommand{\om}{\omega}
\newcommand{\pa}{\partial}
\newcommand{\+}{\dagger}
\newcommand{\fr}{\frac}
\newcommand{\bc}{\begin{center}}
\newcommand{\ec}{\end{center}}
\newcommand{\Ga}{\Gamma}
\newcommand{\de}{\delta}
\newcommand{\De}{\Delta}
\newcommand{\ep}{\epsilon}
\newcommand{\varep}{\varepsilon}
\newcommand{\ka}{\kappa}
\newcommand{\La}{\Lambda}
\newcommand{\si}{\sigma}
\newcommand{\Si}{\Sigma}
\newcommand{\ta}{\tau}
\newcommand{\up}{\upsilon}
\newcommand{\Up}{\Upsilon}
\newcommand{\ze}{\zeta}
\newcommand{\ps}{\psi}
\newcommand{\Ps}{\Psi}
\newcommand{\ph}{\phi}
\newcommand{\vph}{\varphi}
\newcommand{\Ph}{\Phi}
\newcommand{\Om}{\Omega}
\newcommand{\AdrHEPC}{Phenikaa Institute for Advanced Study, Phenikaa University, Nguyen Trac, Duong Noi, Hanoi 100000, Vietnam}
\newcommand{\AdrH}{Institute of Physics, Vietnam Academy of Science and Technology, 10 Dao Tan, Giang Vo, Hanoi 100000, Vietnam}

\title{Physical implications of a double right-handed gauge symmetry} 

\author{Duong Van Loi}
\email{loi.duongvan@phenikaa-uni.edu.vn (corresponding author)}
\affiliation{\AdrHEPC} 
\author{A. E. C\'arcamo Hern\'andez}
\email{antonio.carcamo@usm.cl}
\affiliation{{Universidad T\'ecnica Federico Santa Mar\'{\i}%
		a, Casilla 110-V, Valpara\'{\i}so, Chile}}
\affiliation{{Centro Cient\'{\i}fico-Tecnol\'ogico de Valpara\'{\i}so, Casilla 110-V,
		Valpara\'{\i}so, Chile}}
\affiliation{{Millennium Institute for Subatomic Physics at High-Energy Frontier
		(SAPHIR), Fern\'andez Concha 700, Santiago, Chile}}
\author{N. T. Duy}
\email{ntduy@iop.vast.vn}
\affiliation{\AdrH} 
\author{D. T. Binh}
\email{dinhthanhbinh@tdtu.edu.vn}
\affiliation{Laboratory of Advanced Materials and Natural Resources, Institute for
	Advanced Study in Technology, Ton Duc Thang University, Ho Chi Minh City,
	Vietnam}
\affiliation{Faculty of Applied Sciences, Ton Duc Thang University, Ho Chi Minh City,
	Vietnam}
\author{Cao H. Nam}
\email{caohoangnam@duytan.edu.vn}
\affiliation{Institute of Theoretical and Applied Research, Duy Tan University, Hanoi 100000, Vietnam}
\affiliation{School of Engineering and Technology, Duy Tan University, Da Nang 550000, Vietnam}
\date{\today}

\begin{abstract}
Guided by the flipping principle, we propose a novel extension of the Standard Model based on a double right-handed $U(1)$ gauge symmetry. In this framework, all left-handed fermions are neutral, while right-handed fermions of the third generation carry charges distinct from those of the first two generations. This structure naturally explains the observed  Standard Model fermion mass hierarchy: the heavy masses of the third generation are generated at tree level, while the lighter masses of the first and second generations arise radiatively at the one-loop level. For the active neutrino sector, the tiny masses are generated through a combination of tree-level and two-loop seesaw mechanisms. Crucially, this approach successfully reproduces the observed neutrino mass hierarchy, with the atmospheric mass-squared difference generated at tree level and the solar neutrino mass squared difference emerging at the two-loop level. These hierarchical patterns stem from the interplay between gauge invariance and a residual parity symmetry that survives the spontaneous breaking of the extended gauge group. The same residual symmetry stabilizes a viable scalar singlet dark matter candidate, which we show can reproduce the observed relic abundance while remaining consistent with current direct detection bounds. After addressing constraints from electroweak precision tests and flavor-changing neutral currents, we explore the discovery prospects for the new neutral bosons at existing and future colliders, including the LEP, LHC, and a future ILC.

\end{abstract}

\maketitle

\section{Introduction}
Despite its remarkable successes, the Standard Model (SM) of particle physics remains incomplete, as it fails to account for several fundamental puzzles. These include the pronounced hierarchy in charged fermion masses~\cite{ParticleDataGroup:2024cfk}, the tiny but nonzero masses and mixings of neutrinos~\cite{Kajita:2016cak, McDonald:2016ixn}, and the existence of dark matter (DM) in the universe~\cite{Planck:2018vyg}. A wide range of extensions to the SM have been proposed to address these open questions. Among them, particularly compelling are those that offer a unified explanation of multiple phenomena while remaining testable at current or future experiments. In this work, we propose such a model, guided by the flipping principle.

It is well known that the matter content of the V-A theory, including left-handed fermions, transforms universally as isodoublets under the $SU(2)_L$ symmetry of weak isospin $T_{1,2,3}$~\cite{Feynman:1958ty, Sudarshan:1958vf, Sakurai:1958zz}. The electric charges of these multiplets are given by $Q = \text{diag}(0, -1)$ for $l_{aL} = (\nu_{aL}, e_{aL})^T$ and $Q = \text{diag}(2/3, -1/3)$ for $q_{aL} = (u_{aL}, d_{aL})^T$, where $a = 1,2,3$ denotes the generation index. Since $Q$ neither commutes with nor closes algebraically with the weak isospin generators, i.e., $[Q, T_1 \pm i T_2] = \pm (T_1 \pm i T_2) \neq 0$ and $\mathrm{Tr} Q \neq 0$, the $SU(2)_L$ symmetry must be extended (or flipped) to $SU(2)_L \otimes U(1)_Y$, where $Y = Q - T_3$ is identified as hypercharge. This leads to the electroweak theory, which necessarily includes right-handed fermion singlets $e_{aR}$, $u_{aR}$, and $d_{aR}$ to ensure cancellation of $U(1)_Y$ anomalies~\cite{Glashow:1961tr, Weinberg:1967tq}.

By introducing a dark charge $\mathsf{D}$ as a generalized variant of the electric charge $Q$, gauge invariance and anomaly cancellation require $\mathsf{D} = \text{diag}(\delta, \delta - 1)$ for lepton doublets $l_{aL}$ and $\mathsf{D} = \text{diag}(2/3 - \delta/3,-1/3 - \delta/3)$ for quark doublets $q_{aL}$~\cite{VanDong:2020cjf, VanLoi:2020kdk, VanLoi:2021dzv, VanDong:2021hhg}. For $\delta = 0$, the dark charge $\mathsf{D}$ reproduces the electric charge $Q$, while for $\delta \neq 0$ it becomes a distinct quantum number. Similar to $Q$, the charge $\mathsf{D}$ neither commutes with nor closes algebraically with the weak isospin generators. Consequently, the electroweak sector must be extended to $SU(2)_L \otimes U(1)_Y \otimes U(1)_\mathsf{N}$, where $\mathsf{N} = \mathsf{D} - T_3$ is identified as the hyperdark charge. The presence of this additional Abelian symmetry necessitates three right-handed neutrinos $\nu_{aR}$ to cancel $U(1)_\mathsf{N}$--related anomalies. In addition, a scalar singlet with $\mathsf{D} = -2\delta$ is required to spontaneously break $U(1)_\mathsf{N}$ to a residual parity and to generate large Majorana masses for the right-handed neutrinos.\footnote{In contrast to this construction, in many existing DM models the stabilizing symmetry—typically a discrete $\mathbb{Z}_2$ parity—is imposed by hand at the outset, rather than emerging as a residual symmetry of a spontaneously broken gauge symmetry~\cite{Krauss:2002px,Ma:2006km,LopezHonorez:2006gr,Okada:2010wd,Das:2019pua,Nam:2020byw,Das:2022oyx,KA:2023dyz,VanLoi:2023pkt}.}

A remarkable feature of the model is that the SM Higgs doublet $H = (H^+, H^0)^T$ transforms nontrivially under the dark charge. Its components carry dark charges $\mathsf{D}(H^+)=1$ and $\mathsf{D}(H^0)=0$, ensuring that the electroweak vacuum expectation value (VEV) does not break the dark symmetry. An even more interesting phenomenon emerges at the specific value $\delta = 1/2$. At this point, the hyperdark charge undergoes a structural simplification: all left-handed fermions become automatically neutral, while all nonzero charges reside solely in the right-handed sector. This chiral reorganization has a well-defined theoretical origin. When $\delta = 1/2$, the hyperdark charge precisely matches the $T_{3R}$ quantum number in left–right symmetric models and reproduces Abelian charge patterns arising in decompositions of the grand unified groups $SO(10)$ and $E_6$. Consequently, $\delta = 1/2$ singles out a theoretically preferred Abelian extension---denoted $U(1)_\mathsf{R}$---that possesses a minimal and anomaly-free charge structure naturally aligned with well-motivated ultraviolet completions.

However, fermion generations remain universal under the extended gauge symmetry, as in the SM. Consequently, the observed hierarchical structure of fermion masses cannot be explained without postulating a large hierarchy among Yukawa coupling constants. This motivates us to further flip the hypercharge symmetry $U(1)_Y$ into $U(1)_\mathcal{Y} \otimes U(1)_\mathcal{R}$, such that only right-handed fermions of the first two generations carry nonzero $U(1)_\mathcal{R}$ charges, while remaining fermion multiplets and the SM Higgs doublet remain neutral under $U(1)_\mathcal{R}$, following an approach similar to that proposed in Ref.~\cite{Nomura:2017ezy}.

In this setup, the SM Higgs doublet provides only the tree-level mass terms for the third generation of SM charged fermions. To generate the masses of the first and second fermion families, we introduce an additional scalar doublet charged under the new gauge group $U(1)_\mathcal{R}$, whose VEV is much smaller than that of the SM Higgs. This small VEV is arranged to vanish at tree level by a suitable choice of parameters in the scalar potential and is subsequently induced at the one-loop level after spontaneous breaking of $U(1)_\mathcal{Y} \otimes U(1)_\mathcal{R}$. The loop diagram responsible for this effect involves three scalar fields that are odd under a residual parity. Mechanisms for generating such a small VEV through radiative effects in multi-scalar extensions of the SM have been studied in the literature~\cite{Chang:1986bp,Kanemura:2012rj,Kanemura:2013qva,Nomura:2017ezy}.

Furthermore, tiny masses for active neutrinos are generated from a combination of tree-level and two-loop seesaw mechanisms. Consequently, this mechanism naturally yields the observed neutrino mass hierarchy: the atmospheric mass-squared difference originates at tree level, whereas the solar neutrino mass squared difference emerges at the two-loop level. Finally, the lightest among parity-odd scalar fields running in the loop is stabilized by the residual parity, thereby providing a viable DM candidate compatible with constraints arising from DM relic abundance and DM direct detection.

The rest of this paper is organized as follows. In Sec.~\ref{model}, we present the structure of the model and examine mass spectra in scalar and gauge sectors. Fermion mass generation is analyzed in Sec.~\ref{fermion}. New physics (NP) phenomena and experimental constraints are discussed in Sec.~\ref{phenomena}. In Sec.~\ref{darkmatter}, we study the DM relic abundance and prospects for direct detection. Finally, Sec.~\ref{conclusion} is devoted to our summary and conclusions. Details of the tiny VEV for the additional Higgs doublet and the mass matrix in the scalar sector are collected in Appendices~\ref{poten} and~\ref{matrix}, respectively.

\section{\label{model}The model}
\subsection{Gauge symmetry, particle content, and symmetry breaking}
We consider an extended $2+1$ Higgs doublet model where the scalar content is augmented by the inclusion of several gauge singlet scalar fields and the SM fermion sector is enlarged by the inclusion of right handed Majorana neutrinos. 
Our model is based on the gauge symmetry group
\be
SU(3)_C \otimes SU(2)_L \otimes U(1)_\mathcal{Y} \otimes U(1)_\mathcal{R} \otimes U(1)_\mathsf{R}, \label{gauge}
\ee
where the first two factors are exactly those of the SM gauge symmetry, while the next two arise from a decomposition of the SM hypercharge group, i.e., $U(1)_Y \to U(1)_\mathcal{Y} \otimes U(1)_\mathcal{R}$. The final factor, $U(1)_\mathsf{R}$, represents an alternative hypercharge symmetry motivated by the flipping principle, which introduces a dark charge $\mathsf{D}$ as a variant of the electric charge $Q$~\cite{VanDong:2020cjf, VanLoi:2020kdk, VanLoi:2021dzv, VanDong:2021hhg}. Accordingly, the electric charge and dark charge operators are defined as follows:
\be Q=T_3 + \mathcal{Y} + \mathcal{R}, \hs \mathsf{D}=T_3 + \mathsf{R}.\ee

The fermion content and their charge assignments under the gauge symmetry in Eq.~(\ref{gauge}) are summarized in Table~\ref{tab1}, where the parameter $z$ is an arbitrary nonzero constant, and $\nu_{aR}$ ($a = 1,2$) denote new fermions introduced to ensure anomaly cancellation within each fermion generation. It is worth noting that only right-handed fermions are charged under $U(1)_\mathsf{R}$, whereas only the right-handed fermions of the first two generations carry nonzero charges under $U(1)_\mathcal{R}$.

\begin{table}[h]
\bc
\begin{tabular}{l|cccccccc}
\hline\hline
Multiplets & $SU(3)_C\otimes SU(2)_L$ & $U(1)_\mathcal{Y}$ & $U(1)_\mathcal{R}$ & $U(1)_\mathsf{R}$ & $P_\mathsf{D}$ \\ \hline 
$l_{aL}=(\nu_{aL},e_{aL})^T$ & $({\bf 1}, {\bf 2})$ & $-1/2$ & $0$ & $0$ & $+$\\
$\nu_{\al R}$ & $({\bf 1}, {\bf 1})$ & $-z$ & $z$ & $1/2$ & $+$\\
$\nu_{3R}$ & $({\bf 1}, {\bf 1})$ & $0$ & $0$ & $1/2$ & $+$\\
$e_{\al R}$ & $({\bf 1}, {\bf 1})$ & $-1+z$ & $-z$ & $-1/2$ & $+$\\
$e_{3R}$ & $({\bf 1}, {\bf 1})$ & $-1$ & $0$ & $-1/2$ & $+$\\
$q_{aL}=(u_{aL},d_{aL})^T$ & $({\bf 3}, {\bf 2})$ & $1/6$ & $0$ & $0$ & $+$\\
$u_{\al R}$ & $({\bf 3}, {\bf 1})$ & $2/3-z$ & $z$ & $1/2$ & $+$\\
$u_{3R}$ & $({\bf 3}, {\bf 1})$ & $2/3$ & $0$ & $1/2$ & $+$\\
$d_{\al R}$ & $({\bf 3}, {\bf 1})$ & $-1/3+z$ & $-z$ & $-1/2$ & $+$\\
$d_{3R}$ & $({\bf 3}, {\bf 1})$ & $-1/3$ & $0$ & $-1/2$ & $+$\\
\hline\hline
\end{tabular}
\caption[]{\label{tab1}Fermion content of the model, where $a = 1,2,3$ and $\alpha = 1,2$ are generation indices.}
\ec
\end{table}
In addition to the fermion content listed in Table~\ref{tab1}, the model includes a scalar sector summarized in Table~\ref{tab2}. The scalar doublet $\Phi_1$ is identified as the SM Higgs doublet, responsible for generating large masses for the third-generation of SM charged fermions: the top quark, bottom quark, and tau lepton. The neutral component of the second doublet, $\Phi_2$, acquires a vacuum expectation value (VEV) much smaller than the VEV of the neutral part of $\Phi_1$.  
The second scalar doublet $\Phi_2$ is introduced to generate small masses for the first generation of SM charged fermions. 
The scalar singlets $\chi_1$ and $\chi_3$ are required to generate Majorana masses for $\nu_{\alpha R}$ and $\nu_{3R}$, respectively. Finally, the singlets $\chi_2$ and $\eta_{1,2}$, together with the scalar doublet $\phi$, are included to induce the small VEV for $\Phi_2$ through one-loop radiative corrections.

\begin{table}[h]
\bc
\begin{tabular}{l|cccccccc}
\hline\hline
Multiplets & $SU(3)_C\otimes SU(2)_L$ & $U(1)_\mathcal{Y}$ & $U(1)_\mathcal{R}$ & $U(1)_\mathsf{R}$ & $P_\mathsf{D}$ \\ \hline  
$\Ph_1=(\Ph_1^+,\Ph_1^0)^T$ & $({\bf 1}, {\bf 2})$ & $1/2$ & $0$ & $1/2$ & $+$\\
$\Ph_2=(\Ph_2^+,\Ph_2^0)^T$ & $({\bf 1}, {\bf 2})$ & $1/2-z$ & $z$ & $1/2$ & $+$\\
$\chi_1$ & $({\bf 1}, {\bf 1})$  & $-2z$ & $2z$ & $-1$ & $+$\\
$\chi_2$ & $({\bf 1}, {\bf 1})$  & $-z/3$ & $z/3$ & $0$ & $+$\\
$\chi_3$ & $({\bf 1}, {\bf 1})$  & $0$ & $0$ & $-1$ & $+$\\
$\ph=(\ph^+,\ph^0)^T$ & $({\bf 1}, {\bf 2})$ & $1/2-z/3$ & $z/3$ & $0$ & $-$\\
$\eta_1$ & $({\bf 1}, {\bf 1})$  & $-z/3$ & $z/3$ & $1/2$ & $-$\\
$\eta_2$ & $({\bf 1}, {\bf 1})$  & $0$ & $0$ & $1/2$ & $-$\\
\hline\hline
\end{tabular}
\caption[]{\label{tab2}Scalar content of the model.}
\ec
\end{table}
The remaining gauge symmetry is spontaneously broken in three stages:
\bc \begin{tabular}{c} $SU(2)_L\otimes U(1)_\mathcal{Y}\otimes U(1)_\mathcal{R}\otimes U(1)_\mathsf{R}$ \\
$\downarrow\La_3$\\
$SU(2)_L\otimes U(1)_\mathcal{Y}\otimes U(1)_\mathcal{R}\otimes \mathsf{P_R}$ \\
$\downarrow\La_{1,2}$\\
$SU(2)_L\otimes U(1)_Y\otimes \mathsf{P_R}$\\
$\downarrow v_{1,2}$\\
$U(1)_Q\otimes P_\mathsf{D}$ \end{tabular}\ec
The VEVs responsible for triggering the spontaneous breaking of the gauge symmetry are given by:
\bea \langle\chi_1\rangle &=&\frac{1}{\sqrt2}\La_1,\hs
 \langle\chi_2\rangle =\frac{1}{\sqrt2}\La_2, \hs \langle\chi_3\rangle =\frac{1}{\sqrt2}\La_3,\\
 \langle \Phi_1\rangle &=&\frac{1}{\sqrt2}\begin{pmatrix} 0\\v_1 \end{pmatrix},\hs \langle \Phi_2\rangle =\frac{1}{\sqrt2}\begin{pmatrix} 0\\v_2 \end{pmatrix}.\label{vevs}
 \eea
These VEVs satisfy the hierarchy $\Lambda_{1,2,3} \gg v_{1,2}$, with $v_1^2 + v_2^2 = v^2 \simeq (246~\mathrm{GeV})^2$, to ensure consistency with the SM. In addition, we impose two further hierarchies: $\Lambda_3 \gg \Lambda_{1,2}$ to support the generation of tiny neutrino masses, and $v_2 \ll v_1$ to account for the hierarchical structure of the SM charged fermion masses. The latter is naturally realized by assuming that the VEV of $\Phi_2$ vanishes at tree level and is generated only at one loop, after the second stage of gauge symmetry breaking, as discussed in Appendix~\ref{poten}.
 
After the first stage of gauge symmetry breaking, the group $U(1)_\mathsf{R}$ is reduced to a residual parity $\mathsf{P_R} = (-1)^{2\mathsf{R}}$, as the VEV of $\chi_3$ preserves this symmetry, i.e., $\mathsf{P_R} \langle \chi_3 \rangle = \langle \chi_3 \rangle$. This residual parity remains conserved through the second stage of symmetry breaking. However, in the third stage, $\mathsf{P_R}$ is effectively replaced by a new residual parity $\mathsf{P_D} = (-1)^{2\mathsf{D}}$, as a result of the VEVs of the scalar doublets $\Phi_{1,2}$. This occurs because $\mathsf{R} \langle \Phi_{1,2} \rangle = (0~v_{1,2}/2\sqrt{2})^T \neq 0$ and $\mathsf{P_D} \langle \Phi_{1,2} \rangle = \langle \Phi_{1,2} \rangle$. Moreover, the singlet fields $\chi_{1,2,3}$ also preserve this new parity, satisfying $\mathsf{P_D} \langle \chi_{1,2,3} \rangle = \langle \chi_{1,2,3} \rangle$.

It is well known that spin parity, defined as $P_s = (-1)^{2s}$, is always conserved due to Lorentz invariance. We thus define a final residual parity by combining spin parity with the dark parity as
\be
P_\mathsf{D} = (-1)^{2\mathsf{D} + 2s},
\ee
which remains conserved in our model. This parity plays a similar role to the matter parity in supersymmetric theories~\cite{Martin:1997ns}. The $P_\mathsf{D}$ assignments of all fermions and scalars are shown in the last columns of Tables~\ref{tab1} and~\ref{tab2}, respectively. All fermions, as well as the scalars $\Phi_{1,2}$ and $\chi_{1,2,3}$, are $P_\mathsf{D}$-even, while the scalars $\phi$ and $\eta_{1,2}$ are $P_\mathsf{D}$-odd. As a result, the latter fields cannot acquire VEVs.

\subsection{\label{scalargauge}Scalar sector}

Given the scalar content described above, the scalar potential of our model can be decomposed into three parts, $V = V_1 + V_2 + V_3$, as presented in Appendix~\ref{poten}. The term $V_1$ contains the self-interactions of the fields $\Phi_{1,2}$ and $\chi_{1,2}$. The term $V_2$ includes the self-interactions of $\phi$ and $\eta_{1,2}$, as well as their interactions with $\Phi_{1,2}$ and $\chi_{1,2}$. Finally, $V_3$ consists of the potential for $\chi_3$ and its interactions with the remaining scalar fields. We assume that the parameters of the scalar potential are chosen such that the potential is bounded from below and gives rise to the desired vacuum structure. In particular, we consider a strong hierarchy $|\mu_8| \gg |\mu_{1,2,\dots,7}|$, which ensures that $\chi_3$ effectively decouples from the low-energy spectrum. As implied by $V_3$, the scalar field $\chi_3$ acquires a large VEV, given approximately by $\Lambda_3 \simeq -\mu_8^2 / \lambda_{32}$. We expand $\chi_3$ as
$ \chi_3 = \frac{1}{\sqrt{2}} \left( \Lambda_3 + H_\mathsf{R} + i G_{Z_\mathsf{R}} \right),$
where $H_\mathsf{R}$ is a heavy physical Higgs boson with mass $m_{H_\mathsf{R}} \simeq \sqrt{2} \Lambda_3$, and $G_{Z_\mathsf{R}}$ is the would-be Goldstone boson that is absorbed by the $U(1)_\mathsf{R}$ gauge boson.

Below the $\Lambda_3$ scale, the effective scalar potential takes the form
\be \mathsf{V}_\text{eff} \simeq V_1 + V_2 + \left( \mu_0^2 \, \Phi_1^\dagger \Phi_2 + \mathrm{H.c.} \right),\ee
which generates the scalar mass matrices presented in Appendix~\ref{matrix}.  Assuming the hierarchy $\Lambda_1 \sim \Lambda_2 \gg v_1 \sim\sqrt{-\mu_0^2}\gg v_2$, the mass-squared matrix $M_S^2$ of the $P_\mathsf{D}$-even CP-even neutral scalars $(S_1, S_2, S_3, S_4)$ can be diagonalized using the seesaw approximation, separating the light state $S_1$ from the heavy states $S_2$, $S_3$, and $S_4$. In the resulting physical basis $(H, \mathcal{H}, H_1, H_2)$, the mass eigenstates are approximately given by
\begin{align}
H &\simeq S_1 - \epsilon_1 S_2 - \epsilon_2 S_3 - \epsilon_3 S_4, \\
\mathcal{H} &\simeq \epsilon_1 S_1 + S_2, \hs H_1 \simeq \epsilon_2 S_1 + S_3, \hs H_2 \simeq \epsilon_3 S_1 + S_4,
\end{align}
where the mixing parameters $\epsilon_{1,2,3}$ are small due to hierarchical suppression, i.e.,
\be
\epsilon_1 \simeq -\frac{2 \lambda_1 v_1 v_2}{\mu_0^2}, \hs
\epsilon_2 \simeq \frac{\lambda_7 \lambda_8 v_1}{(\lambda_7^2 - 4 \lambda_5 \lambda_6) \Lambda_2}, \hs
\epsilon_3 \simeq \frac{2 \lambda_5 \lambda_8 \Lambda_1 v_1}{(4 \lambda_5 \lambda_6 - \lambda_7^2) \Lambda_2^2}.
\ee

The lightest state $H$ is decoupled from the heavy sector and plays the role of the SM-like Higgs boson, with mass approximately given by \be
m_H^2 \simeq 2 \lambda_1 v_1^2 - \frac{\mu_0^2 v_2}{v_1} - \epsilon_1 \mu_0^2 - \epsilon_2 \lambda_8 \Lambda_1 v_1 - \epsilon_3 \lambda_{10} \Lambda_2 v_1, \ee
which lies at the electroweak scale. The state $\mathcal{H}$, with mass
\be
m_\mathcal{H}^2 \simeq -\frac{\mu_0^2 v_1}{v_2},
\ee
can be safely treated as a separate heavy eigenstate, since its mixing with $H_{1,2}$ is suppressed by $v_2 / \Lambda_{1,2}$. The remaining scalars $H_1$ and $H_2$ mix through a $2 \times 2$ submatrix, approximately given by the bottom-right block of $M_S^2$. Diagonalizing this submatrix yields two physical states, $\mathcal{H}_1$ and $\mathcal{H}_2$, with large masses at the $\Lambda_{1,2}$ scales.

For the $P_\mathsf{D}$-even CP-odd neutral scalars $(A_1, A_2, A_3, A_4)$, we obtain two massless eigenstates,
\be G_{Z_\mathcal{R}} = \fr{6\La_1 A_3+\La_2 A_4}{\sqrt{36\La_1^2+\La_2^2}}, \hs \mathcal{G} = \fr{\La_2 A_3-6\La_1 A_4}{\sqrt{36\La_1^2+\La_2^2}},\ee
in addition to two eigenstates arising from the $M_A^2$ matrix,
\be G_Z = c_\varphi A_1 + s_\varphi A_2, \hs \mathcal{A} = s_\varphi A_1 - c_\varphi A_2,
\ee
where $G_Z$ is also massless, while $\mathcal{A}$ is a heavy physical state with mass
\be
m_\mathcal{A}^2 = -\frac{\mu_0^2 v^2}{v_1 v_2}=m_\mathcal{H}^2\frac{v^2}{v_1^2}.
\ee
Above, we define $c_\varphi \equiv \cos\varphi = v_1 / v$ and $s_\varphi \equiv \sin\varphi = v_2 / v$ for brevity. The massless modes $G_{Z_\mathcal{R}}$ and $G_Z$ are identified as the Goldstone bosons absorbed by the $Z_\mathcal{R}$ and $Z$ gauge bosons, respectively. The state $\mathcal{G}$, on the other hand, remains a physical massless boson. As discussed in Ref.~\cite{VanLoi:2023utt}, the presence of such a massless scalar does not pose a phenomenological problem. The reason is that $\mathcal{G}$ does not couple directly to SM particles, except for the SM-like Higgs boson $H$. Even in that case, the coupling strength is controlled by the potential parameters $\lambda_{8,10}$, which are required to be $|\lambda_{8,10}| \lesssim \mathcal{O}(10^{-3})$ to suppress observable effects. Furthermore, $\mathcal{G}$ decouples from the thermal bath in the early universe at a temperature of order $\mathcal{O}(1)$~GeV, provided that all new scalar states reside at the TeV scale. Consequently, it remains consistent with standard cosmology.

Since the masslessness of $\mathcal{G}$ originates from an accidental global symmetry, it may be lifted by Planck-suppressed operators. A gauge-invariant dimension-five operator consistent with the charge assignment in Table~\ref{tab2} is $\frac{1}{M_\mathrm{P}} (\Phi_2^\dagger \Phi_1) \chi_2^3+\mathrm{H.c.}$, which explicitly breaks the accidental symmetry, where $M_\mathrm{P}$ denotes the Planck mass~\cite{Akhmedov:1992hi, Rothstein:1992rh}. For typical scales $v_1 \sim 10^2$~GeV, $v_2 \sim 1$~GeV, and $\Lambda_2 \sim 10$~TeV, this operator induces $\Delta V\sim v_1v_2\La_2^3/M_P$ and yields a pseudo-goldstone mass of order $m_\mathcal{G}\sim 10^2$~eV. Such a mass is small enough to evade stringent bounds from tests of non-Newtonian forces~\cite{Dupays:2006dp}.

For the remaining neutral scalar fields, the model contains two $3 \times 3$ mass-squared matrices, $\mathcal{M}_S^2$ for the CP-even states $(S_5, S_6, S_7)$ and $\mathcal{M}_A^2$ for the CP-odd states $(A_5, A_6, A_7)$ as shown in Eqs. (\ref{MS}) and (\ref{MA}). In both matrices, the off-diagonal elements are strongly suppressed by the hierarchy $v_2\sim\mu\ll v_1\ll\La_{1,2}$, and are therefore much smaller than the diagonal entries. As a result, mixing effects can be safely neglected, and the fields $S_{5,6,7}$ and $A_{5,6,7}$ may be treated as approximate mass eigenstates. In this limit, each pair ($S_i,A_i$) ($i=5,6,7$) becomes mass-degenerate and naturally combines into a physical complex scalar field,
\be \phi^0 \equiv \frac{1}{\sqrt2} (S_5+iA_5),\hs \eta_1 \equiv \frac{1}{\sqrt2}(S_6+iA_6),\hs \eta_2 \equiv \frac{1}{\sqrt2}(S_7+iA_7), \ee
with approximate masses
\be m_{\phi^0}^2 \simeq \mu_5^2+\fr 1 2 (\la_{20}\La_1^2+\la_{21}\La_2^2),\hs 
m_{\eta_1}^2 \simeq \mu_6^2+\fr 1 2 (\la_{26}\La_1^2+\la_{27}\La_2^2),\hs m_{\eta_2}^2\simeq \mu_7^2+\fr 1 2 (\la_{30}\La_1^2+\la_{31}\La_2^2). \ee

For the charged scalar sector, the field $\phi^\pm$ is a physical eigenstate with a large mass given by
\be 
m^2_{\phi^\pm} \simeq \mu_5^2 + \frac{1}{2} \left(\lambda_{20} \Lambda_1^2 + \lambda_{21} \Lambda_2^2 \right),
\ee
which lies at the $\Lambda_{1,2}$ scale. The charged components of the Higgs doublets, $\Phi_1^\pm$ and $\Phi_2^\pm$, mix through a $2 \times 2$ submatrix identical to the top-left block of the full charged scalar mass matrix $M_C^2$. Diagonalizing this submatrix yields a massless eigenstate $G_W^\pm = c_\varphi \, \Phi_1^\pm + s_\varphi \, \Phi_2^\pm$, which is identified as the Goldstone boson absorbed by the $W^\pm$ gauge boson, and a heavy charged scalar
\be
\mathcal{H}^\pm = s_\varphi \, \Phi_1^\pm - c_\varphi \, \Phi_2^\pm,
\ee
with mass
\be
m^2_{\mathcal{H}^\pm} = -\frac{(2 \mu_0^2 + \lambda_4 v_1 v_2) v^2}{2 v_1 v_2}=m_\mathcal{A}^2-\frac{\la_4}{2}v^2,
\ee
which typically lies at the TeV scale.

\subsection{Gauge boson sector}

The covariant derivative in our model is defined as
\[
D_\mu = \partial_\mu + i g_s t_p G_{p\mu} + i g T_j A_{j\mu} + i g_1 \mathcal{Y} B_{1\mu} + i g_2 \mathcal{R} B_{2\mu} + i g_\mathsf{D} \mathsf{R} C_\mu,
\]
where $(g_s, g, g_1, g_2, g_\mathsf{D})$ denote the gauge coupling constants, $(t_p, T_j, \mathcal{Y}, \mathcal{R}, \mathsf{R})$ are the corresponding generators, and $(G_{p\mu}, A_{j\mu}, B_{1\mu}, B_{2\mu}, C_\mu)$ are the gauge bosons associated with the groups $SU(3)_C$, $SU(2)_L$, $U(1)_\mathcal{Y}$, $U(1)_\mathcal{R}$, and $U(1)_\mathsf{R}$, respectively. As previously mentioned, the gauge boson of the $U(1)_\mathsf{R}$ group, denoted by $Z_\mathsf{R} \equiv C_\mu$, acquires a large mass through the spontaneous breaking of the symmetry at the $\Lambda_3$ scale. Its mass is approximately given by
\be
m_{Z_\mathsf{R}} \simeq 2 g_\mathsf{D} \Lambda_3,
\ee
rendering it effectively decoupled from the low-energy particle spectrum. In this context, any possible kinetic mixing between $C_\mu$ and the $U(1)_\mathcal{Y}$ and $U(1)_\mathcal{R}$ gauge bosons, $B_{1\mu}$ and $B_{2\mu}$, becomes irrelevant and can be safely neglected. Moreover, potential kinetic mixing between $B_{1\mu}$ and $B_{2\mu}$, if present, would induce effects that are subdominant compared to the mass mixing effects arising from spontaneous symmetry breaking. These effects are therefore suppressed, in agreement with the results of Refs.~\cite{VanLoi:2023kgl, VanLoi:2024ptt}.

Besides the heavy gauge boson $Z_\mathsf{R}$, the masses of the remaining gauge bosons arise from the scalar kinetic terms, $\sum_S (D^\mu S)^\dagger (D_\mu S)$, where the sum runs over all scalar fields $S = \Phi_{1,2}, \chi_{1,2}$. Since all scalar multiplets are color singlets, the $SU(3)_C$ symmetry remains unbroken and the gluons remain massless, as expected. Defining the charged gauge bosons as
\be
W^\pm_\mu \equiv \frac{1}{\sqrt{2}} (A_{1\mu} \mp i A_{2\mu}),
\ee
we obtain the mass term $\mathcal{L} \supset \frac{g^2 v^2}{4} W^{\mu+} W^-_\mu$,
which is identical to the SM result. The $W$ gauge boson is a physical eigenstate by itself with mass
\[
m_W^2 = \frac{g^2 v^2}{4},
\]
which lies at the electroweak scale. This matches the SM prediction and allows us to identify $v \simeq 246$~GeV.

The neutral gauge bosons $A_{3\mu}$, $B_{1\mu}$, and $B_{2\mu}$ mix through the mass matrix $M_0^2$, 
\be
\mathcal{L} \supset \frac{1}{2} (A^\mu_3,\, B_1^\mu,\, B_2^\mu)\, M_0^2 \, (A_{3\mu},\, B_{1\mu},\, B_{2\mu})^T,
\ee
where
\be M_0^2=\fr{g^2}{4}\left(\begin{array}{ccc} v^2 & \frac{g_1(2zv_2^2-v^2)}{g} & -\frac{2g_2zv_2^2}{g} \\
\frac{g_1(2zv_2^2-v^2)}{g} & \frac{g_1^2[4z^2(36\La_1^2+\La_2^2)+9v_1^2+9(1-2z)^2v_2^2]}{9g^2} & \frac{2g_1g_2z[9v_2^2(1-2z)-2z(36\La_1^2+\La_2^2)]}{9g^2} \\
-\frac{2g_2zv_2^2}{g} & \frac{2g_1g_2z[9v_2^2(1-2z)-2z(36\La_1^2+\La_2^2)]}{9g^2} & \frac{4g_2^2z^2(36\La_1^2+\La_2^2+9v_2^2)}{9g^2}\end{array}\right). \ee 
The neutral gauge boson mass matrix $M_0^2$ possesses an exact zero eigenvalue, corresponding to the massless photon. To identify the physical mass eigenstates, we define the Weinberg angle $\theta_W$ and a mixing angle $\theta$ as
\[
s_W \equiv \sin\theta_W = \frac{g_1 g_2}{\sqrt{g^2(g_1^2 + g_2^2) + g_1^2 g_2^2}}, \qquad
s_\theta \equiv \sin\theta = \frac{g_2}{\sqrt{g_1^2 + g_2^2}} = \frac{g_Y}{g_1},
\]
where $g_Y$ is the effective hypercharge coupling constant. Using these definitions, the gauge boson eigenstates can be expressed as orthonormal linear combinations of the original fields $A_{3\mu}$, $B_{1\mu}$, and $B_{2\mu}$:
\begin{align}
A_\mu &= s_W\, A_{3\mu} + c_W (s_\theta\, B_{1\mu} + c_\theta\, B_{2\mu}), \label{eq:photon} \\
Z_\mu &= c_W\, A_{3\mu} - s_W (s_\theta\, B_{1\mu} + c_\theta\, B_{2\mu}), \label{eq:Zboson} \\
Z'_\mu &= c_\theta\, B_{1\mu} - s_\theta\, B_{2\mu}, \label{eq:Zprime}
\end{align}
where $c_W \equiv \cos\theta_W$ and $c_\theta \equiv \cos\theta$. $A_\mu$ is the photon field, $Z_\mu$ is identified with the SM $Z$ boson, and $Z'_\mu$ is a new heavy neutral gauge boson that is orthogonal to both $A_\mu$ and $Z_\mu$.

In the basis $(A_\mu, Z_\mu, Z'_\mu)$, the photon field $A_\mu$ remains massless and fully decoupled, as expected. However, the neutral gauge bosons $Z_\mu$ and $Z'_\mu$ mix through a $2 \times 2$ submatrix of the full mass matrix. The relevant entries of this submatrix are given by
\bea m^2_Z &=& \fr{g^2v^2}{4c^2_W}, \hs m^2_{ZZ'}= \fr{g^2v^2[4zs_\varphi^2-1-c_{2\theta})]t_W}{4c_Ws_{2\theta}},\\
m^2_{Z'} &=& \frac{g^2\{9v^2[3-4(1-2z)(2z-c_{2\theta})+c_{4\theta}+8z(1-2z+c_{2\theta})c_{2\varphi}]+32z^2(36\La_1^2+\La_2^2)\}t_W^2}{72 s_{2\theta}^2}.\eea
Diagonalizing this submatrix yields two physical neutral gauge bosons,
\be Z_{1\mu} = c_\varepsilon Z_\mu - s_\varepsilon Z'_\mu,\hs Z_{2\mu} = s_\varepsilon Z_\mu + c_\varepsilon Z'_\mu,\ee
with the corresponding masses approximately given by
\be
m_{Z_1}^2 \simeq m_Z^2 - \frac{m_{ZZ'}^4}{m_{Z'}^2}, \hs
m_{Z_2}^2 \simeq m_{Z'}^2,\label{gaugemasses}
\ee
under the assumption $v \ll \Lambda_{1,2}$, which ensures that $Z_1$ closely resembles the SM $Z$ boson. The mixing angle $\varepsilon$ between $Z$ and $Z'$ is also small and approximately given by
\be t_{2\varepsilon}\simeq -\fr{9v^2c_\theta^3s_\theta}{2s_W z^2 (36\La_1^2+\La_2^2)},\ee
which confirms that the mixing is suppressed by the ratio $v^2 / \Lambda^2_{1,2}$. As a result, \( Z_1 \) can be safely identified as the observed SM-like \( Z \) boson, while \( Z_2 \) is a heavy neutral gauge boson with mass at the \(\mathcal{O}(\Lambda_{1,2})\) scale.

\subsection{Fermion-gauge boson interactions}

This subsection presents the interactions between fermions and gauge bosons, which originate from the kinetic term of fermion fields: $\mathcal{L}_\text{kin}\supset\sum_F \bar{F} i\gamma^\mu D_\mu F$, where the sum runs over all fermion multiplets in the model. The covariant derivative is rewritten in terms of physical gauge bosons as
\bea D_\mu &=& \pa_\mu + ig_s t_p G_{p\mu} + i gs_W Q A_\mu + i g (T_+W^+_\mu + \mathrm{H.c.})  \crn
&&+ \fr{ig}{c_W}\left[c_\varep (T_3-s^2_W Q)+s_\varep \fr{s_W}{s_\theta c_\theta}(\mathcal{R}-c^2_\theta Y)\right]Z_{1\mu} \crn
&&+ \fr{ig}{c_W}\left[s_\varep (T_3-s^2_W Q)-c_\varep \fr{s_W}{s_\theta c_\theta}(\mathcal{R}-c^2_\theta Y)\right]Z_{2\mu}+ ig_\mathsf{D}\mathsf{R}Z_{\mathsf{R}\mu},\label{eq:covariant_derivative}\eea 
where $T_\pm = (T_1 \pm i T_2)/\sqrt{2}$ are the weight-raising and lowering operators of the $SU(2)_L$ group. From Eq.~\eqref{eq:covariant_derivative}, it is clear that the interactions of fermions with gluons, the photon, and the \( W^\pm \) bosons remain unchanged compared to the SM:
\be \mathcal{L} \supset -\fr{g_s}{2} \bar{q} \ga^\mu \la_p q G_{p\mu}-eQ(f')\bar{f}'\ga^\mu f' A_\mu - \fr{g}{\sqrt{2}} [(\bar{\nu}_{aL} \ga^\mu e_{aL} + \bar{u}_{aL}\ga^\mu d_{aL})W^+_\mu + \mathrm{H.c.}],\ee
where $\lambda_p$ ($p = 1, \dots, 8$) are the Gell-Mann matrices. The fields $q$ and $f'$ denote quarks and charged fermions in mass eigenstate basis respectively, while $\nu_{aL}$, $e_{aL}$, $u_{aL}$, and $d_{aL}$ (with $a = 1,2,3$) represent the fermion components in the gauge eigenbasis.

The interactions between the neutral gauge bosons $Z_{1,2}$ and fermions can be written in the following form:
\be \mathcal{L} \supset -\fr{g}{2c_W}\left[C_L^{Z_I}\bar{\nu}_{aL}\gamma^\mu\nu_{aL} + C_R^{Z_I}\bar{\nu}_{\al R}\gamma^\mu\nu_{\al R} + \bar{f}\ga^\mu\left(g^{Z_I}_V(f)- g^{Z_I}_A(f)\ga_5\right)f \right]Z_{I\mu},\ee
where $I=1,2$, while $\nu_{aL}$, $\nu_{\al R}$, and $f$ denote the left-handed neutrinos, right-handed neutrinos, and charged fermions in the gauge basis, respectively. Additionally, $a = 1,2,3$ and $\al = 1,2$ are the family indices as previously defined. The coupling coefficients $C_{L,R}^{Z_I}$ and $g_{V,A}^{Z_I}(f)$ are given by the following expressions:
\bea 
C_L^{Z_1} &=& c_\varep + s_\varep\fr{s_W}{t_\theta},\hs C_R^{Z_1} = s_\varep\fr{2zs_W}{s_\theta c_\theta},\\
g_V^{Z_1}(f) &=& c_\varep\left(T_3(f_L)-2Q(f)s_W^2\right)+s_\varep\fr{s_W}{s_\theta c_\theta} \left[\mathcal{R}(f_L)+\mathcal{R}(f_R)-c^2_\theta(Y(f_L)+Y(f_R))\right],\\
g_A^{Z_1}(f) &=& c_\varep T_3(f_L)+s_\varep\fr{s_W}{s_\theta c_\theta}\left[\mathcal{R}(f_L)-\mathcal{R}(f_R)-c^2_\theta(Y(f_L)-Y(f_R))\right],\\
C_{L,R}^{Z_2} &=& C_{L,R}^{Z_1}|_{c_\varep\to s_\varep, s_\varep\to -c_\varep},\hs g_{V,A}^{Z_2}(f) = g_{V,A}^{Z_1}(f)|_{c_\varep\to s_\varep, s_\varep\to -c_\varep}.
\eea 
For completeness, we present the vector and axial-vector couplings of the $Z_1$ ($Z_2$) boson to the charged fermions in Table~\ref{tab3} (Table~\ref{tab4}). It is evident that the couplings of the $Z_1$ boson to the SM fermions reduce to those of the SM $Z$ boson in the limit $\varep\to 0$.

\begin{table}[h]
\bc
\begin{tabular}{c|ccc}
\hline\hline
$f$ & $g^{Z_1}_V(f)$ && $g^{Z_1}_A(f)$ \\
\hline 
$e_\al$ & $c_\varep\left(2s^2_W-\fr 1 2\right)+s_\varep s_W\left(\fr{3}{2t_\theta}-\fr{z}{s_\theta c_\theta}\right)$ && $-\fr 1 2 c_\varep - s_\varep s_W\left(\frac{1}{2t_\theta}-\frac{z}{s_\theta c_\theta}\right)$\\
$e_3$ & $c_\varep\left(2s^2_W-\fr 1 2\right)+s_\varep\fr{3s_W}{2t_\theta}$ && $-\fr 1 2 c_\varep - s_\varep\frac{s_W}{2t_\theta}$\\
$u_\al$ & $c_\varep\left(\fr 1 2 -\fr 4 3 s^2_W\right)-s_\varep s_W\left(\fr{5}{6t_\theta}-\fr{z}{s_\theta c_\theta}\right)$ && $\fr 1 2 c_\varep + s_\varep s_W\left(\frac{1}{2t_\theta}-\frac{z}{s_\theta c_\theta}\right)$\\
$u_3$ & $c_\varep\left(\fr 1 2 -\fr 4 3 s^2_W\right)-s_\varep\fr{5s_W}{6t_\theta}$ && $\fr 1 2 c_\varep + s_\varep\frac{s_W}{2t_\theta}$\\
$d_\al$ & $c_\varep\left(\fr 2 3 s^2_W-\fr 1 2\right)+s_\varep s_W\left(\fr{1}{6t_\theta}-\fr{z}{s_\theta c_\theta}\right)$ && $-\fr 1 2 c_\varep - s_\varep s_W\left(\frac{1}{2t_\theta}-\frac{z}{s_\theta c_\theta}\right)$\\
$d_3$ & $c_\varep\left(\fr 2 3 s^2_W-\fr 1 2\right)+s_\varep\fr{s_W}{6t_\theta}$ && $-\fr 1 2 c_\varep - s_\varep\frac{s_W}{2t_\theta}$\\
\hline\hline
\end{tabular}
\caption[]{\label{tab3}Vector and axial-vector couplings of the $Z_1$ boson to charged fermions.}
\ec
\end{table}
\begin{table}[h]
\bc
\begin{tabular}{c|ccc}
\hline\hline
$f$ & $g^{Z_2}_V(f)$ && $g^{Z_2}_A(f)$ \\
\hline 
$e_\al$ & $s_\varep\left(2s^2_W-\fr 1 2\right)-c_\varep s_W\left(\fr{3}{2t_\theta}-\fr{z}{s_\theta c_\theta}\right)$ && $-\fr 1 2 s_\varep + c_\varep s_W\left(\frac{1}{2t_\theta}-\frac{z}{s_\theta c_\theta}\right)$\\
$e_3$ & $s_\varep\left(2s^2_W-\fr 1 2\right)-c_\varep\fr{3s_W}{2t_\theta}$ && $-\fr 1 2 s_\varep + c_\varep\frac{s_W}{2t_\theta}$\\
$u_\al$ & $s_\varep\left(\fr 1 2 -\fr 4 3 s^2_W\right)+ c_\varep s_W\left(\fr{5}{6t_\theta}-\fr{z}{s_\theta c_\theta}\right)$ && $\fr 1 2 s_\varep - c_\varep s_W\left(\frac{1}{2t_\theta}-\frac{z}{s_\theta c_\theta}\right)$\\
$u_3$ & $s_\varep\left(\fr 1 2 -\fr 4 3 s^2_W\right)+ c_\varep\fr{5s_W}{6t_\theta}$ && $\fr 1 2 s_\varep - c_\varep\frac{s_W}{2t_\theta}$\\
$d_\al$ & $s_\varep\left(\fr 2 3 s^2_W-\fr 1 2\right)- c_\varep s_W\left(\fr{1}{6t_\theta}-\fr{z}{s_\theta c_\theta}\right)$ && $-\fr 1 2 s_\varep + c_\varep s_W\left(\frac{1}{2t_\theta}-\frac{z}{s_\theta c_\theta}\right)$\\
$d_3$ & $s_\varep\left(\fr 2 3 s^2_W-\fr 1 2\right)- c_\varep\fr{s_W}{6t_\theta}$ && $-\fr 1 2 s_\varep + c_\varep\frac{s_W}{2t_\theta}$\\
\hline\hline
\end{tabular}
\caption[]{\label{tab4}Vector and axial-vector couplings of the $Z_2$ boson to charged fermions.}
\ec
\end{table}

It is noteworthy that the charges $Q$, $T_3$, $Y$, and $\mathsf{R}$ are universal across all flavors of neutrinos, charged leptons, up-type quarks, and down-type quarks, whereas the charge $\mathcal{R}$ is flavor-dependent. As a consequence, both $Z_1$ and $Z_2$ can mediate flavor-changing interactions among fermions. The flavor violation induced by $Z_1$ originates from the $Z$--$Z'$ mixing and is therefore strongly suppressed. In contrast, the flavor-changing effects mediated by $Z_2$ can be sizable, even when $\varep = 0$. In the following analysis, we focus on the flavor-changing effects arising from $Z_2$ and disregard those due to $Z_1$, given that $|\varep| \lesssim 10^{-3}$.

\section{\label{fermion}Fermion mass and mixing}
\begin{figure}[h]
\centering
\includegraphics[scale=1]{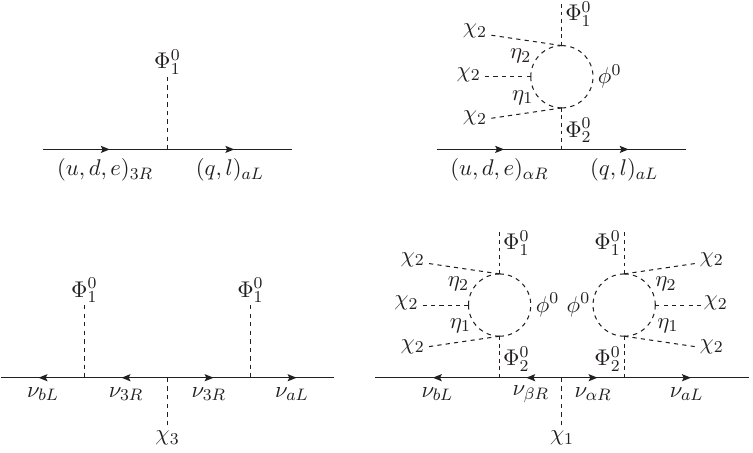}
\caption[]{\label{fig1}Feynman diagrams contributing to the entries of the mass matrices for SM charged fermions (first row) and light active neutrinos (second row). Here $a,b=1,2,3$ and $\alpha,\beta=1,2$.}
\end{figure}

With the particle content and symmetries specified in Tables \ref{tab1} and \ref{tab2}, the following renormalizable Yukawa interactions arise
\bea \mathcal{L}_\text{Yuk} &=&  h^\nu_{a\al}\bar{l}_{aL}i\sigma_2 \Phi^*_2\nu_{\al R} + h^e_{a\al}\bar{l}_{aL}\Phi_2 e_{\al R} + h^\nu_{a3}\bar{l}_{aL}i\sigma_2 \Phi^*_1\nu_{3R} + h^e_{a3}\bar{l}_{aL}\Phi_1 e_{3R} \crn
&&+ h^u_{a\al}\bar{q}_{aL}i\sigma_2 \Phi^*_2u_{\al R} + h^d_{a\al}\bar{q}_{aL}\Phi_2 d_{\al R} + h^u_{a3}\bar{q}_{aL}i\sigma_2 \Phi^*_1u_{3R} + h^d_{a3}\bar{q}_{aL}\Phi_1 d_{3R}\crn
&& + \fr 1 2 f^\nu_{\al\bet}\bar{\nu}_{\al R}^c\chi_1\nu_{\bet R} + \fr 1 2 f^\nu_{33}\bar{\nu}_{3R}^c\chi_3\nu_{3R} + \mathrm{H.c.,}\label{Lyuk}\eea
where $\sigma_2$ is the second Pauli matrix, and all Yukawa couplings $h$'s and $f$'s are dimensionless. After the scalar fields acquire their VEVs at both tree and one-loop levels, the above interactions generate mass matrices for up-type quarks, down-type quarks, and charged leptons, all sharing the same texture:
\be M = -\frac{1}{\sqrt2}\begin{pmatrix} h_{11}v_2 & h_{12}v_2 & h_{13}v_1\\
h_{21}v_2 & h_{22}v_2 & h_{23}v_1\\
h_{31}v_2 & h_{32}v_2 & h_{33}v_1 \end{pmatrix},\label{chmassma} \ee
where the superscripts on the Yukawa couplings are omitted for brevity and should be understood from the context. This mass generation mechanism for charged fermions is illustrated by the diagrams in the first row of Fig.~\ref{fig1}.

To derive the physical masses of charged fermions and the Cabibbo–Kobayashi–Maskawa (CKM) matrix, we decompose the Hermitian product $MM^\dagger \equiv \mathcal{M}$ into two parts: $\mathcal{M} = \mathcal{M}^\text{tree} + \mathcal{M}^\text{loop}$, where $\mathcal{M}^\text{tree}$ and $\mathcal{M}^\text{loop}$ contain terms proportional to $v_1^2$ and $v_2^2$, respectively. Given that $v_1 \gg v_2$, the leading contributions of $\mathcal{M}$ arise from $\mathcal{M}^\text{tree}$, yielding
\be \mathcal{V}_L^\dag \mathcal{M}^\text{tree}\mathcal{V}_L=\text{diag}\left(0,0,\fr 1 2 (h_{13}^2+h_{23}^2+h_{33}^2)v_1^2\right), \ee
which corresponds to the masses of the third-family charged fermions, i.e., 
\be m_3^2 = \frac{1}{2}(h_{13}^2 + h_{23}^2 + h_{33}^2)v_1^2.\ee
The unitary matrix $\mathcal{V}_L$ takes the form
\be \mathcal{V}_L=\begin{pmatrix}
\frac{1}{\sqrt{2+(t_{13}+t_{23})^2}} & \frac{1+t_{23}(t_{13}+t_{23})}{\sqrt{[2+(t_{13}+t_{23})^2](1+t_{13}^2+t_{23}^2)}} & \frac{t_{13}}{\sqrt{1+t_{13}^2+t_{23}^2}}\\
\frac{1}{\sqrt{2+(t_{13}+t_{23})^2}} & -\frac{1+t_{13}(t_{13}+t_{23})}{\sqrt{[2+(t_{13}+t_{23})^2](1+t_{13}^2+t_{23}^2)}} & \frac{t_{23}}{\sqrt{1+t_{13}^2+t_{23}^2}}\\
-\frac{t_{13}+t_{23}}{\sqrt{2+(t_{13}+t_{23})^2}} & \frac{t_{23}-t_{13}}{\sqrt{[2+(t_{13}+t_{23})^2](1+t_{13}^2+t_{23}^2)}} & \frac{1}{\sqrt{1+t_{13}^2+t_{23}^2}}
\end{pmatrix}, \ee
where we have conveniently denoted $t_{13}=h_{13}/h_{33}$ and $t_{23}=h_{23}/h_{33}$. 

To find the next-to-leading-order contributions of $\mathcal{M}$, we transform it to a new basis via the rotation $\mathcal{M}'=\mathcal{V}_L^\dag\mathcal{M}\mathcal{V}_L$. In this rotated basis, the resulting matrix $\mathcal{M}'$ features a dominant $3\times 3$ element proportional to $v_1^2$, while all remaining entries are suppressed by $v_2^2$. This hierarchical structure allows us to apply the seesaw approximation to isolate the light (upper-left) block of $\mathcal{M}'$, defined as 
\be \delta m^2\equiv\begin{pmatrix}
\delta m_{11}^2 & \delta m_{12}^2 \\ \delta m_{21}^2 & \delta m_{22}^2
\end{pmatrix} = \begin{pmatrix}
(\mathcal{V}_L^\dag)_{1i}(\mathcal{M})_{ij}(\mathcal{V}_L)_{j1} & (\mathcal{V}_L^\dag)_{1i}(\mathcal{M})_{ij}(\mathcal{V}_L)_{j2} \\ (\mathcal{V}_L^\dag)_{2i}(\mathcal{M})_{ij}(\mathcal{V}_L)_{j1} & (\mathcal{V}_L^\dag)_{2i}(\mathcal{M})_{ij}(\mathcal{V}_L)_{j2}
\end{pmatrix},\ee 
for $i,j=1,2,3$. Neglecting the seesaw-induced mixing, which is suppressed by the ratio $v_2^2/v_1^2$, the submatrix $\delta m^2$ yields two nonzero eigenvalues corresponding to the physical masses of the charged fermions in the first and second families,
\be m_{1,2}^2=\fr 1 2 \left[\delta m^2_{11}+\delta m_{22}^2\mp \sqrt{(\delta m^2_{11}-\delta m^2_{22})^2+4\delta m^4_{21}}\right], \ee
which are proportional to $v_2^2$, and a rotation angle $\xi$ to be 
\be t_{2\xi}=\fr{2\delta m^2_{21}}{\delta m_{22}^2-\delta m_{11}^2}.\ee

Hence, the gauge eigenstates and mass eigenstates are related through the matrix $V_L=\mathcal{V}_L\mathcal{V}_\xi$, where $\mathcal{V}_\xi=\{\{c_\xi,s_\xi,0\},\{-s_\xi,c_\xi,0\},\{0,0,1\}\}$. For each sector of the charged fermions, the transformation is explicitly given by 
\be(u_1~u_2~u_3)_L^T=V_{uL}(u~c~t)_L^T, \hs (d_1~d_2~d_3)_L^T=V_{dL}(d~s~b)_L^T, \hs (e_1~e_2~e_3)_L^T=V_{eL}(e~\mu~\tau)_L^T,\ee 
where the superscripts on $V$ correspond to those on the Yukawa couplings. The CKM matrix is then defined by $V_\mathrm{CKM}=V_{uL}^\dag V_{dL}$. 

Concerning the neutral leptons, the neutrinos $\nu_{aL}$ and $\nu_{aR}$ acquire a Dirac mass matrix $M_D$, which has the same structure as the matrix $M$ in Eq. (\ref{chmassma}), but with the Yukawa couplings $h^\nu_{ab}$. In addition, the right-handed neutrinos $\nu_{aR}$ receive a Majorana mass matrix of the form
\be M_M = -\frac{1}{\sqrt2}\begin{pmatrix} f^\nu_{11}\La_1 & f^\nu_{12}\La_1 & 0\\
f^\nu_{21}\La_1 & f^\nu_{22}\La_1 & 0\\
0 & 0 & f^\nu_{33}\La_3 \end{pmatrix}. \ee
The corresponding Feynman diagrams are shown in the second row of Fig.~\ref{fig1}. Since $\La_{1,3}\gg v_{1,2}$, the canonical seesaw mechanism is naturally implemented, leading to tiny masses for the active neutrinos. These masses are generated both at tree level and through two-loop radiative corrections. 

At tree level, the Dirac mass matrix receives contributions solely from $v_1$. Consequently, the seesaw mechanism yields a light active neutrino mass matrix of the form
\be (m_\nu^\text{tree})_{ab}=\fr{h^\nu_{a3}h^\nu_{3b}}{f_{33}^\nu}\frac{v_1^2}{\sqrt2 \La_3}, \ee
which is of rank 1. This implies that only one neutrino acquires mass at tree level, which is incompatible with experimental observations~\cite{ParticleDataGroup:2024cfk}. At loop level, the Dirac mass matrix is generated from $v_2$, while
$v_1$ is absent. The corresponding radiative seesaw contribution to the neutrino mass matrix becomes
\be (m_\nu^\text{loop})_{ab}=-\left(M_DM_M^{-1}M_D^T\right)_{ab}|_{v_1\to 0}\sim \fr{(h^\nu_{a\al})^2}{f^\nu_{\al\bet}}\frac{v_2^2}{\La_1}. \ee
This loop-induced mass matrix has rank 2, thereby generating masses for two additional neutrinos and bringing the model into agreement with experimental data~\cite{ParticleDataGroup:2024cfk}. Assuming representative parameter values such as $v_1\sim 10^2$ GeV, $v_2\sim 1$ GeV, $h^\nu_{ab}\sim 10^{-3}$, and $f^\nu_{\al\bet,33}\sim 1$, the observed neutrino mass scale $m_\nu\sim 0.1$ eV implies that $\La_1\sim 10$ TeV and $\La_3\sim 10^6$ TeV. In an alternative scenario where $h^\nu_{ab}\sim f^\nu_{33}\sim 1$, reproducing the correct neutrino mass scale would require significantly higher values for the symmetry-breaking scales, namely $\La_1\sim 10^7$ TeV and $\La_3\sim 10^{11}$ TeV.

The total neutrino mass matrix is given by
\be (m_\nu)_{ab}=(m_\nu^\text{tree})_{ab}+ (m_\nu^\text{loop})_{ab},\ee
which has full rank (rank 3) and thus yields three non-zero neutrino masses, in agreement with experimental observations~\cite{ParticleDataGroup:2024cfk}. Specifically, the mass eigenvalues $m_{1,2,3}$ are obtained through the diagonalization \be (V^\nu_L)^T m_\nu V^\nu_L=\text{diag}(m_{\nu_e},m_{\nu_\mu},m_{\nu_\tau}),\ee
where $V^\nu_L$ is a unitary matrix that relates the flavor (physical) neutrino states $\nu'_L=(\nu_{e L}~\nu_{\mu L}~\nu_{\tau L})^T$ to the gauge eigenstates $\nu_L=(\nu_{1L}~\nu_{2L}~\nu_{3L})^T$ via $\nu_L=V_{\nu L}\nu'_L$. Accordingly, the Pontecorvo-Maki-Nakagawa-Sakata (PMNS) matrix is defined as $U_\mathrm{PMNS}=V_{\nu L}^\dag V_{e L}$. 
It is worth emphasizing that the model naturally reproduces the observed neutrino mass hierarchy, with the atmospheric and solar neutrino mass-squared differences arising at tree and two-loop levels, respectively. This hierarchical pattern is a direct consequence of the separation between the effective scales $\La_1$ and $\La_3$,, associated with the atmospheric and solar sectors, together with the flavor structure of the corresponding Yukawa couplings. Finally, the model predicts two heavy neutrino states, which are combinations of the right-handed neutrinos $\nu_{\al R}$, with masses at the scale $\La_1$, and an additional very heavy neutrino, predominantly $\sim \nu_{3R}$, with a mass at the scale $\La_3$.

As summarized in Table~\ref{tabfermion}, the model under consideration successfully reproduces the observed masses of the SM charged fermions, the neutrino mass-squared differences, the mixing parameters and the CP phases. 

The eigenvalue problem associated with the mass matrices for SM charged fermons and light active neutrinos is solved, determining a parameter set consistent with the experimental values of the physical observables of the quark and lepton sectors. One of the 
benchmark solutions corresponds to $v_2\simeq 0.52$ GeV, $\La_1=10$ TeV, and $\La_3=10^6$ TeV, and employs the following structures for the fermion mass matrices:
\bea  M_u &\simeq&\left(
\begin{array}{ccc}
 -0.199985 + 0.15933 i & -0.0155587 - 0.162743 i & 3.41237 - 1.2098 i \\
 -0.357103 + 0.282796 i & -0.0295511 - 0.29138 i & 7.91538 - 1.30834 i \\
 (1.231 - 0.089 i)\times 10^{-6} & (1.180 + 0.678 i)\times 10^{-6} & 168.889 - 35.964 i \\
\end{array}
\right)\mathrm{GeV},\label{Mu}\\
 M_d &\simeq& \left(
\begin{array}{ccc}
 0.00920894 & 0.0347929 & 0.0043808 \\
 0.00621922 & 0.0392079 & 0.0259326 \\
 0.0010463 & 0.0166569 & 2.85991 \\
\end{array}
\right)\mathrm{GeV},\label{Md}\\
 M_e &\simeq& \left(
\begin{array}{ccc}
 0.259686 & -0.4745 & 0.000693803 \\
 102.704 & 5.77402 & 0.000513704 \\
 0.000750844 & 0.00164944 & 1747.43 \\
\end{array}
\right)\mathrm{MeV},\\
m_\nu &\simeq& \left(
\begin{array}{ccc}
 3.83781 + 0.984023 i & -1.40753 - 2.86153 i & -6.59195 - 3.03522 i \\
 -1.40753 - 2.86153 i & 26.8893 - 0.296409 i & 21.3523 - 0.0438046 i \\
 -6.59195 - 3.03522 i & 21.3523 - 0.0438046 i & 28.548 + 0.241133 i \\
\end{array}
\right)\mathrm{meV}. \eea 

\begin{table}[
]
\begin{center}
\begin{tabular}{l c c|l c c}
\hline\hline
Observable & Experimental value & Model value & Observable & Experimental value & Model value \\ \hline
$m_{u}$ [MeV] &  $1.24(22)$ & $1.2428$ & $m_e$ [MeV] & $0.4883266(17)$ & 0.488327\\ 
$m_{c}$ [GeV] & $0.62(2)$ & $0.619993$ & $m_\mu$ [MeV] & $102.87267(21)$ & 102.866\\ 
$m_{t}$ [GeV] & $172.9(4)$ & $172.899$ & $m_\tau$ [MeV] & $1747.43(12)$ & 1747.43\\ 
$m_{d}$ [MeV] & $2.69(19)$ & $2.69183$ & $\Delta m_{21}^{2}$ $[10^{-5}$eV$^{2}]$ & $7.49(19)$ & 7.49003 \\ 
$m_{s}$ [MeV] & $53.5(4.6)$ & $53.3857$ & $\Delta m_{31}^{2}$ $[10^{-3}$eV$^{2}]$ & $2.513_{-0.019}^{+0.021}$ & 2.513\\ 
$m_{b}$ [GeV] & $2.86(3)$ & $2.86008$ & $\sin^2\theta^{(l)}_{12}/10^{-1}$ & $3.08_{-0.11}^{+0.12}$ & 3.07999 \\ 
$\sin \theta^{(q)} _{12}/10^{-1}$ & $2.2501(68)$ & $2.2501$ & $\sin^2\theta^{(l)}_{23}/10^{-1}$ & $4.70_{-0.13}^{+0.17}$ & 4.70076\\ 
$\sin \theta^{(q)} _{23}/10^{-2}$ & $4.183_{-0.069}^{+0.079}$ & $4.18315$ & $\sin^2\theta^{(l)}_{13}/10^{-2}$ & $2.215^{+0.056}_{-0.058}$ & 2.215\\ 
$\sin \theta^{(q)} _{13}/10^{-3}$ & $3.732^{+0.090}_{-0.085}$ & $3.73191$ & $\delta^{(l)}_{\mathrm{CP}}$ $[^\circ]$ & $212_{-41}^{+26}$ & 211.926\\ 
$J_q/10^{-5}$ & $3.12^{+0.13}_{-0.12}$ & $3.11966$ & & & \\ \hline\hline
\end{tabular}
\end{center}
\caption{Experimental values of the SM charged fermion masses, neutrino mass-squared differences, and mixing parameters~\cite{ParticleDataGroup:2024cfk, Xing:2020ijf, Esteban:2024eli}, along with the corresponding model predictions obtained from the best-fit solution.}
\label{tabfermion}
\end{table}

\section{\label{phenomena}Constraints}
\subsection{Electroweak precision test}
As discussed in subsection~\ref{scalargauge}, the model predicts a tree-level mixing between the SM $Z$ boson and a new neutral gauge boson $Z'$. This mixing leads to a reduction in the physical mass of the $Z$ boson compared to its SM prediction, as shown in Eq.~(\ref{gaugemasses}). Given the precise experimental measurement of the $Z$-boson mass, $m_Z=91.1880\pm 0.0020$ GeV~\cite{ParticleDataGroup:2024cfk}, we impose the constraint $|\Delta m_Z|< 0.0020$ GeV to maintain consistency with observations.

While the $Z$--$Z'$ mixing reduces the physical mass of the SM $Z$ boson, the mass of the $W$ boson remains unchanged. As a result, the parameter $\rho\equiv m^2_W/(c^2_Wm^2_Z)$, which is equal to unity at tree level in the SM, receives a deviation due to the NP contributions arising from the $Z$--$Z'$ mixing. This deviation can be estimated as
\be\Delta\rho = \frac{m^2_W}{c_W^2m_{Z_1}^2}-1\simeq\frac{m_{ZZ'}^4}{m_Z^2m_{Z'}^2}, \ee
where $m_{Z_1}$ is the mass of the lightest neutral gauge boson (identified with the observed SM $Z$ neutral gauge boson), and $m_{ZZ'}$ denotes the off-diagonal mass term arising from the mixing. Based on the global fit to the $\rho$ parameter, $\rho=1.00031\pm 0.00019$~\cite{ParticleDataGroup:2024cfk}, we adopt a conservative upper bound of $\Delta\rho< 0.00050$ to ensure compatibility with electroweak precision constraints.

The $Z$--$Z'$ mixing also modifies the well-measured vector and axial-vector couplings of the SM $Z$ boson to fermions, which are given by $g_V^Z(f)=T_3(f)-2Q(f)s^2_W$ and $g_A^Z(f)=T_3(f)$. As a result of the mixing, these couplings are shifted according to
\be g_{V,A}^{Z_1}(f)= g_{V,A}^Z(f)+\mathcal{O}(\varepsilon),\ee
where $\varepsilon$ being the $Z$--$Z'$ mixing parameter. Electroweak precision data place stringent limits on deviations from the SM predictions, allowing NP effects only if $|\varepsilon|\lesssim 10^{-3}$~\cite{ALEPH:2005ab,Erler:2009jh}. This requirement therefore imposes an additional constraint on the viable model parameter space. 

Another constraint can be derived from the precision measurement of the total decay width of the observed $Z_1$ boson. Experimentally, the width is measured as $\Gamma^{\text{exp}}_{Z_1}=2.4955\pm0.0023$ GeV, while the SM predicts $\Gamma^{\text{SM}}_{Z_1}=2.4942\pm0.0008$ GeV~\cite{ParticleDataGroup:2024cfk}. To remain consistent with observation, we require the shift in decay width to satisfy $|\Delta\Gamma_{Z_1}|<0.0044$ GeV. In our model, the shift in decay width of $Z_1$ can be approximately calculated as~\cite{VanLoi:2025fmy}
\bea
\Delta\Gamma_{Z_1}&\simeq&\frac{m_Z}{6\pi}\left(\frac{g}{2c_W}\right)^2\left\{\sum_{f}N^f_C\left[\left|G_{V}^{Z_1}(f)\right|^2\mathrm{Re}[\delta_V^{f}]+\left|G_{A}^{Z_1}(f)\right|^2\mathrm{Re}[\delta_A^{f}]\right]+2\sum_i\left|G_L^{Z_1}(\nu_{iL})\right|^2\mathrm{Re}[\delta^{\nu_{iL}}]\right\}\crn
&&+\frac{\Delta m_Z}{12\pi}\left(\frac{g}{2c_W}\right)^2\left\{\sum_f N_C^f\left[\left|G_{V}^{Z_1}(f)\right|^2+\left|G_{A}^{Z_1}(f)\right|^2\right]+2\sum_i\left|G_L^{Z_1}(\nu_{iL})\right|^2\right\},
\eea
where $N_C^f$ is the number of color degrees of freedom for the charged fermion $f$, and $G_{V,A}^{Z_1}(f)$, $\delta^f_{V,A}$ denote vector and axial-vector couplings of $Z_1$ to $f$ in the mass basis and their respective fractional shifts. Similarly, $G_L^{Z_1}(\nu_{iL})$ and $\delta^{\nu_{iL}}$ refer to the couplings of $Z_1$ to the physical neutrinos and their corresponding shifts~\cite{VanLoi:2025fmy}.

\begin{figure}[h]
\centering
\includegraphics[scale=0.42]{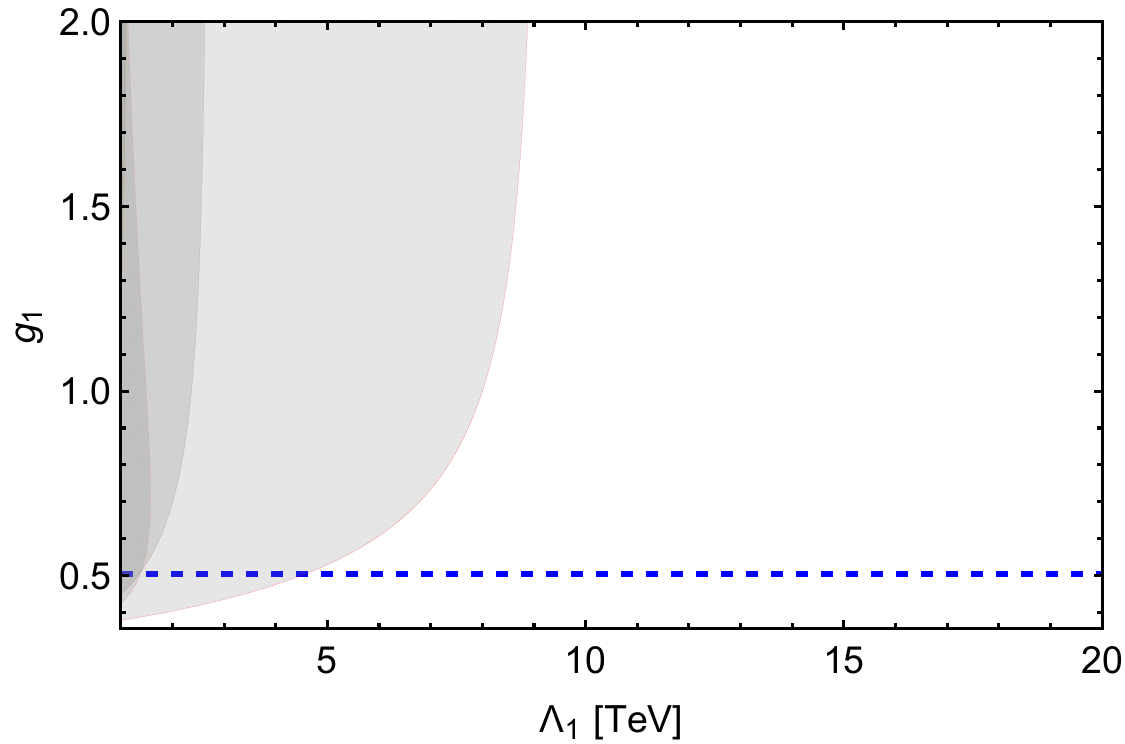}
\includegraphics[scale=0.415]{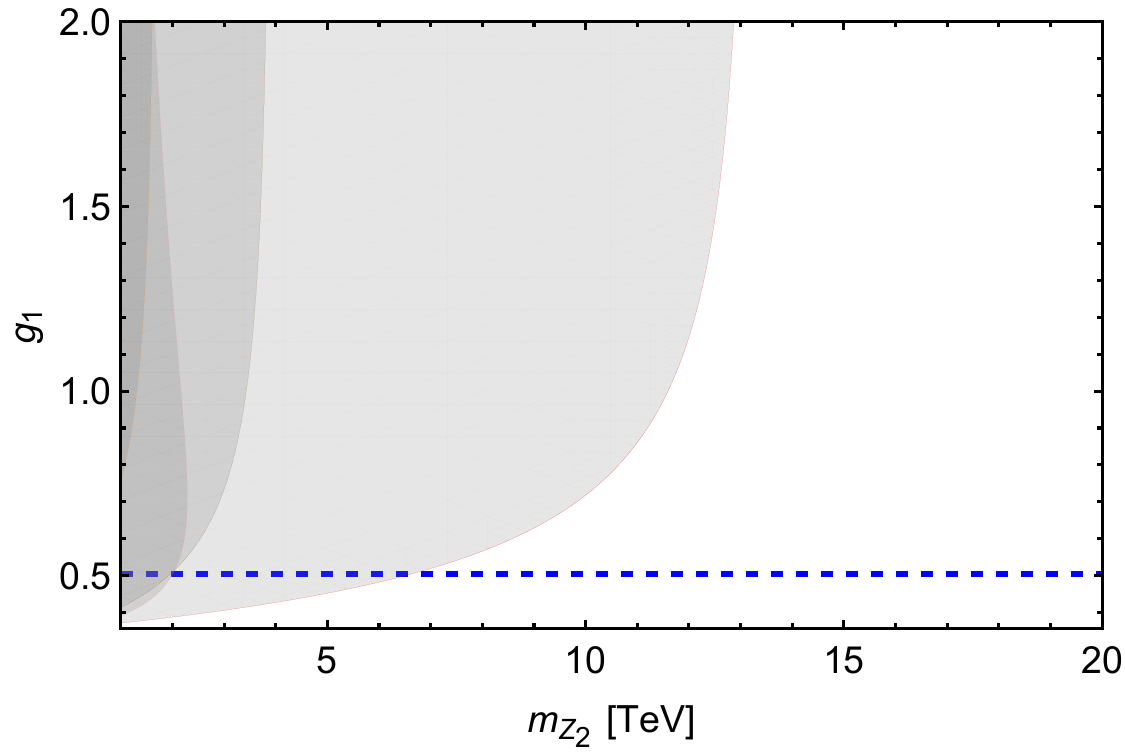}
\caption[]{\label{plotLa1vsg1}The red, brown, pink, and orange curves represent the exclusion bounds derived from the $Z$-boson mass shift~\cite{ParticleDataGroup:2024cfk}, the $\rho$ parameter~\cite{ParticleDataGroup:2024cfk}, the $Z$--$Z'$ mixing parameter $\varepsilon$~\cite{ALEPH:2005ab, Erler:2009jh}, and the total decay width of $Z_1$~\cite{ParticleDataGroup:2024cfk}, respectively. The shaded regions are excluded by current experimental data. The dashed blue line indicates the gauge coupling unification condition $g_1=g_2=\sqrt2 g_Y$.}
\end{figure}

In Fig.~\ref{plotLa1vsg1}, we illustrate the correlation between the coupling $g_1$ of the new $U(1)_\mathcal{Y}$ group and the NP scale $\La_1$ (left panel), as well as the mass of the new gauge boson $m_{Z_2}$ (right panel). In this analysis, we fix $z=1$ and $v_2\simeq 0.52~\mathrm{GeV}$, and assume $\La_1\simeq\La_2$ for simplicity. The relation $g_2=g_1g_Y/\sqrt{g_1^2-g_y^2}$ is also employed. 
The constraint from the $Z$-boson mass shift $\Delta m_Z$ is substantially stronger than those arising from the $\rho$-parameter, the mixing parameter $\varepsilon$, and the decay width shift $\Delta\Gamma_{Z_1}$. Furthermore, under the gauge coupling unification condition $g_1=g_2=\sqrt2 g_Y$, represented by the dashed blue line, the model predicts lower bounds of $\La_1\gtrsim 4.7$ TeV and $m_{Z_2}\gtrsim 6.9$ TeV. We emphasize that these results remain stable under variations of the parameters $z$ and $v_2$ within reasonable ranges, specifically $0.01\lesssim |z|\lesssim 20$ and $0.05 \mathrm{~GeV} \lesssim v_2\lesssim 10 \mathrm{~GeV}$.

\subsection{FCNCs}
Because the right-handed fermions carry non-universal charges under the  $U(1)_{\mathcal{Y}}\otimes U(1)_{\mathcal{R}}$ gauge group, our model naturally generates tree-level Flavor-Changing Neutral Current (FCNC) processes, which serve as a key signature of new physics. These include $\Delta F=2$ observables such as $\bar{B}_s-B_s$ and $\bar{B}_d-B_d$ mixing, as well as rare decays like $B_s \to \ell^+\ell^-$, all mediated by the neutral gauge boson $Z'$. Furthermore, the Yukawa interactions in Eq.~(\ref{Lyuk}) induce flavor-changing scalar interactions via the neutral scalars $H$, $\mathcal{H}$, and $\mathcal{A}$, providing additional contributions to FCNC observables. In this section, we investigate a set of these observables arising from both the $Z'$ boson and the neutral scalar fields.

For the FCNCs induced by the $Z'$ gauge boson, the relevant interactions are given by
\be
\mathcal{L} \supset \bar{q}_{\al R}i\ga^{\mu}(ig_1\mathcal{Y}_{q_{\al R}}B_{1\mu}+ig_2\mathcal{R}_{q_{\al R}}B_{2\mu})q_{\al R}+\bar{q}_{3 R}i\ga^{\mu}(ig_1\mathcal{Y}_{q_{3 R}}B_{1\mu}+ig_2\mathcal{R}_{q_{3R}}B_{2\mu})q_{3R}, \label{fcnc1}
\ee
where $q=u,d$. The gauge fields $B_{1\mu}$ and $B_{2\mu}$ are related to the physical neutral gauge bosons $Z_\mu$ and $Z'_\mu$ through the relations $B_{1\mu}= -s_Ws_{\theta}Z_{\mu}+c_{\theta}Z'_{\mu}$, $B_{2\mu}= -s_{W}c_{\theta}Z_{\mu}-s_{\theta}Z'_{\mu}$. Substituting these relations into Eq.~(\ref{fcnc1}) and transforming to the mass eigenbasis, where $q_{L,R}=V_{qL,R}q'_{L,R}$, we obtain the FCNC interactions,
\be 
\mathcal{L}^{\text{gauge}}_{\text{FCNC}} = \bar{q}'_{iR}[\Ga^{Z'}_{qR}]_{ij}q_{jR}\ga^{\mu}Z'_\mu, \ee
where $[\Ga^{Z'}_{qR}]_{ij}$ denote the flavor-violating couplings induced by the $Z'_\mu$ gauge boson, 
\be
[\Ga^{Z'}_{qR}]_{ij}=\pm(g_1c_{\theta}+g_2s_{\theta})z\sum_{\al=1,2}[V^*_{qR}]_{\al i}[V_{qR}]_{\al j},\label{Gaij}
\ee
and the sign $+(-)$ corresponds to up-type (down-type) quarks. The quark mixing matrices $V_{qL,R}$ are numerically obtained from benchmark points given in Eqs.~(\ref{Mu}) and (\ref{Md}). 

Besides, the model also includes FCNC sources arising from scalar exchanges. These contributions originate from the Yukawa interactions given in Eq.~(\ref{Lyuk}). For the scalar components $\Phi_1^0 \simeq \frac{1}{\sqrt2}[v_1+ H + i(c_\varphi G_Z+s_\varphi \mathcal{A})]$ and $\Phi_2^0 \simeq \frac{1}{\sqrt2}[v_2+ \mathcal{H} + i(s_\varphi G_Z-c_\varphi \mathcal{A})]$, the flavor-changing scalar interactions can be written as
\be 
\mathcal{L}^{\text{scalar}}_{\text{FCNC}}=- \bar{f}'_i [\Ga^{H}_{fR}]_{ij}P_Rf'_jH-\bar{f}'_i [\Ga^{\mathcal{H}}_{fR}]_{ij}P_Rf'_j\mathcal{H}-i\bar{f}'_i [\Ga^{\mathcal{A}}_{fR}]_{ij}P_Rf'_j\mathcal{A} + \mathrm{H.c.},
\ee 
where $P_R=\fr 1 2 (1+\gamma_5)$, $f=u,d,e$, and the corresponding flavor-violating couplings are given by
\bea 
[\Ga^{H}_{fR}]_{ij}&= &\fr{1}{v_1}\sum_{k=1}^3\sum_{a=1}^3[V_{fL}^*]_{ai}[V_{fL}]_{ak}M^f_k[V_{fR}^*]_{3k}[V_{fR}]_{3j},\label{GaSij1}\\\relax
[\Ga^{\mathcal{H}}_{fR}]_{ij}&=&\fr{1}{v_2}\sum_{k=1}^3\sum_{a=1}^3\sum_{\al=1}^2[V_{fL}^*]_{ai}[V_{fL}]_{\al k}M^f_k[V_{fR}^*]_{\al k}[V_{fR}]_{\al j},\\\relax
[\Ga^{\mathcal{A}}_{fR}]_{ij}&= &\fr{\pm g}{2m_W}\sum_{k=1}^3\sum_{a=1}^3\sum_{\al=1}^2\left\{t_{\varphi}[V_{fL}^*]_{ai}[V_{fL}]_{ak}M^f_k[V_{fR}^*]_{3k}[V_{fR}]_{3j}\right.\crn
&&\left.- \fr{1}{t_{\varphi}}[V_{fL}^*]_{ai}[V_{fL}]_{\al k}M^f_k[V_{fR}^*]_{\al k}[V_{fR}]_{\al j}\right\}, \label{GaSij}
\eea 
with the positive (negative) sign corresponding to $f=e,d$ ($f=u$). 

For the $\Delta F=2$ flavor-changing processes, the effective Hamiltonian can be written as 
\bea 
\mathcal{H}^{\Delta F=2}_{\text{eff}}&=&\fr{G_F^2m_W^2(V^*_{tq}V_{tb})^2}{16\pi^2}[C^q_{\text{SM}}(\bar{b}_{L}\ga_{\mu}q_{L})^2+C^{q}_{RR}(\bar{b}_{R}\ga_{\mu}q_{R})^2+\tilde{C}^q_{LL} (\bar{b}_Rq_{L})^2\crn &&+\tilde{C}^q_{RR}(\bar{b}_Lq_R)^2+\tilde{C}^q_{LR}(\bar{b}_Rq_L)(\bar{b}_Lq_R)],\label{Heff}
\eea 
where $\al$ and $\beta$ denote color indices. The Wilson coefficients (WCs) $C^{q}_{\text{SM}}$ represent the SM contribution to the $\bar{B}_q$-$B_q$ meson mixings and are given by~\cite{Buras:2012dp}
\bea 
C^q_{\text{SM}}&=&4S_0(x_t)\eta_{2B},
\eea 
where $S_0(m_i^2/m_W^2)$ is the Inami-Lim function~\cite{Inami:1980fz}, and $\eta_{2B}\simeq 0.84$ accounts for the next-to-leading order (NLO) QCD corrections~\cite{Buras:1990fn}.\footnote{For the $K$ meson, the SM prediction $\Delta m_K^{\text{SM}}$ suffers from large and poorly controlled theoretical uncertainties due to long-distance effects. Consequently, the constraint from $\Delta m_K$ is weaker than those from $\Delta m_{B_{s,d}}$, and it is therefore omitted in the present analysis.} It is worth emphasizing that only the right-handed fermions carry non-universal charges under the gauge group $U(1)_{\mathcal{Y}}\otimes U(1)_{\mathcal{R}}$. As a result, the NP contributions from the additional gauge boson $Z'$ affect solely the WC $C^q_{RR}$. The NP WCs $C^{q}_{RR}$ and $\tilde{C}^q_{LL,RR,LR}$ are defined at the matching scale $\mu=m_{Z'}$: 
\bea  
C^{q}_{RR}&=&\fr{16\pi^2}{G_F^2m_W^2(V^*_{tq}V_{tb})^2}\fr{([\Ga^{Z'}_{dR}]_{3q})^2}{m_{Z'}^2},\label{CqRRZp}\\ 
\tilde{C}^q_{LL}&=&\fr{-16\pi^2}{G_F^2m_W^2(V^*_{tq}V_{tb})^2}\left[\fr{([\Ga^{H}_{dR}]^*_{q3})^2}{m_H^2}+\fr{([\Ga^{\mathcal{H}}_{dR}]^*_{q3})^2}{m_{\mathcal{H}}^2}-\fr{([\Ga^{\mathcal{A}}_{dR}]^*_{q3})^2}{m_{\mathcal{A}}^2}\right],\\ 
\tilde{C}^q_{RR}&=&\fr{-16\pi^2}{G_F^2m_W^2(V^*_{tq}V_{tb})^2}\left[\fr{([\Ga^{H}_{dR}]_{3q})^2}{m_H^2}+\fr{([\Ga^{\mathcal{H}}_{dR}]_{3q})^2}{m_{\mathcal{H}}^2}-\fr{([\Ga^{\mathcal{A}}_{dR}]_{3q})^2}{m_{\mathcal{A}}^2}\right], \\
\tilde{C}^q_{LR}&=&\fr{-32\pi^2}{G_F^2m_W^2(V^*_{tq}V_{tb})^2}\left[\fr{[\Ga^{H}_{dR}]_{3q}[\Ga^{H}_{dR}]^*_{q3}}{m_H^2}+\fr{([\Ga^{\mathcal{H}}_{dR}]_{3q}[\Ga^{\mathcal{H}}_{dR}]^*_{q3}}{m_{\mathcal{H}}^2}+\fr{[\Ga^{\mathcal{A}}_{dR}]_{3q}[\Ga^{\mathcal{A}}_{dR}]^*_{q3}}{m_{\mathcal{A}}^2}\right],
\eea 
where the flavor-violating couplings $[\Ga^{Z'}_{dL,R}]_{ij}$ and $[\Ga^{H,\mathcal{H},\mathcal{A}}_{dL,R}]_{ij}$ are defined in Eq.~(\ref{Gaij}) and Eqs.~(\ref{GaSij1})--(\ref{GaSij}), respectively. With the effective Hamiltonian for $\Delta F=2$ processes given in Eq.~(\ref{Heff}), the ratio between the SM+NP and SM contributions to the meson mass differences can be expressed as  
\bea 
\ep_{B_q}&=&\fr{\Delta m_{B_q}^{\text{SM+NP}} }{\Delta m_{B_q}^{\text{SM}}}=\left|1+\fr{M_{12}^{q,\text{NP}}}{M_{12}^{q,\text{SM}}}\right|\crn 
&=&\left|1+\fr{C^{q}_{RR}}{C^q_{\text{SM}}}\eta_{B_q}^{6/23}+\fr{(\tilde{C}^q_{RR}+\tilde{C}^q_{LL})}{C^q_{\text{SM}}}\left(\fr{m_{B_q}}{m_b+m_q}\right)^2\left(\fr{-5B^{(2)}_{B_q}\eta_{22}+B^{(3)}_{B_q}\eta_{32}}{12B^{(1)}_{B_q}}\right) \right. \crn  
&& +\left. \fr{\tilde{C}^q_{LR}}{C^q_{\text{SM}}}\left[\fr{1}{6}+\left(\fr{m_{B_q}}{m_b+m_q}\right)^2\right]\eta_4\fr{B^{(4)}_{B_q}}{B^{(1)}_{B_q}}\right|.
\eea 
Here, $\eta_{22}=0.983\,\eta_{B_q}^{-0.63}+0.017\,\eta_{B_q}^{0.717}$, $\eta_{32}=-0.064\,\eta_{B_q}^{-0.63}+0.064\,\eta_{B_q}^{0.717}$, and $\eta_4=\eta_{B_q}^{-24/23}$, where the coefficient $\eta_{B_q}=\al_s(\mu_{\text{NP}})/\al_s(\mu_{B_q})$ represents the leading-order (LO) QCD correction obtained via renormalization group evolution (RGE) from the new-physics scale $\mu_{\text{NP}}=m_{\text{NP}}$ down to the hadronic scale $\mu_{B_q}=4.16$ GeV~\cite{Bagger:1997gg}. The impact of NP on the mass difference $\Delta m_{B_q}$ can then be constrained by the 2$\sigma$ experimental ranges given in Ref.~\cite{VanLoi:2024ptt}: 
\bea 
&& \fr{\Delta m_{B_s}^{\text{SM+NP}} }{\Delta m_{B_s}^{\text{SM}}}=\fr{\Delta m_{B_s}^{\text{exp}} }{\Delta m_{B_s}^{\text{SM}}}\in [0.9072,1.0419],\label{Bq_const1}\\ 
&& \fr{\Delta m_{B_d}^{\text{SM+NP}} }{\Delta m_{B_d}^{\text{SM}}}=\fr{\Delta m_{B_d}^{\text{exp}} }{\Delta m_{B_d}^{\text{SM}}}\in [0.8728,1.0222]. \label{Bq_const2}
\eea 

We also emphasize that our model predicts NP contributions to the $\Delta S=1$ rare decay $B_s\to \mu^+\mu^-$ at tree-level. This decay can be described by the effective Hamiltonian
\be 
\mathcal{H}^ {\Delta S=1}_{\text{eff}}=-\fr{4G_FV^*_{ts}V_{tb}}{\sqrt{2}}[C_{10}^{\text{SM}}(\bar{s}_L\ga_{\mu}b_L)(\bar{\mu}\ga^{\mu}\ga_5\mu)+C_{10}^{'}(\bar{s}_R\ga_{\mu}b_R)(\bar{\mu}\ga^{\mu}\ga_5\mu)\sum_{i=S,P}(C_i\mathcal{O}_i+C'_i\mathcal{O}'_i)],
\ee
where the scalar and pseudoscalar operators induced by the neutral scalars $H,\mathcal{H}$ and $\mathcal{A}$ are defined as 
\bea 
&& \mathcal{O}_S=\fr{e^2}{16\pi^2}(\bar{s}P_R b)(\bar{\mu}\mu), \hs  \mathcal{O}'_S=\fr{e^2}{16\pi^2}(\bar{s}P_L b)(\bar{\mu}\mu),\\ 
&& \mathcal{O}_P=\fr{e^2}{16\pi^2}(\bar{s}P_R b)(\bar{\mu}\ga_5\mu), \hs  \mathcal{O}'_P=\fr{e^2}{16\pi^2}(\bar{s}P_L b)(\bar{\mu}\ga_5\mu).
\eea   
Since the FCNCs associated with the new gauge boson $Z'$ only arise from the right-handed sector, there is no NP contribution to the left-handed operator $(\bar{s}_L\ga_{\mu}b_L)(\bar{\mu}\ga^{\mu}\ga_5\mu)$, similar to the case of the $\Delta S=2$ observables. In the SM, the WC $C_{10}^{\text{SM}}$ has been computed up to next-to-next-to-leading order (NNLO) in QCD, with a numerical value of $C_{10}^{\text{SM}}=-4.194$. The non-SM WCs are given by
\bea 
C^{'}_{10}&=&-\fr{16\pi^2}{e^2}\fr{g\sqrt{2}}{8c_WG_FV^*_{ts}V_{tb}}\fr{[\Ga^{Z'}_{dR}]_{23}[\tilde{g}^{Z'}_A]_{22}}{m_{Z'}^2},\label{WCBsmmZp} \\ 
C_{S}&=&\fr{16\pi^2}{e^2}\fr{\sqrt{2}}{4G_FV^*_{ts}V_{tb}}\left[\fr{[\Ga^{H}_{dR}]_{23}[\mathrm{Re}([\Ga^H_{eR}]_{22})}{m_{H}^2}+\fr{[\Ga^{\mathcal{H}}_{dR}]_{23}\mathrm{Re}([\Ga^{\mathcal{H}}_{eR}]_{22})}{m_{\mathcal{H}}^2}-i\fr{[\Ga^{\mathcal{A}}_{dR}]_{23}\mathrm{Im}([\Ga^{\mathcal{A}}_{eR}]_{22})}{m_{\mathcal{A}}^2}\right],\\ 
C'_{S}&=&\fr{16\pi^2}{e^2}\fr{\sqrt{2}}{4G_FV^*_{ts}V_{tb}}\left[\fr{[\Ga^{H}_{dR}]^*_{32}[\mathrm{Re}([\Ga^H_{eR}]_{22})}{m_{H}^2}+\fr{[\Ga^{\mathcal{H}}_{dR}]^*_{32}\mathrm{Re}([\Ga^{\mathcal{H}}_{eR}]_{22})}{m_{\mathcal{H}}^2}+i\fr{[\Ga^{\mathcal{A}}_{dR}]^*_{32}\mathrm{Im}([\Ga^{\mathcal{A}}_{eR}]_{22})}{m_{\mathcal{A}}^2}\right],\\ 
C_{P}&=&-\fr{16\pi^2}{e^2}\fr{\sqrt{2}}{4G_FV^*_{ts}V_{tb}}\left[\fr{[\Ga^{\mathcal{A}}_{dR}]_{23}\mathrm{Re}([\Ga^{\mathcal{A}}_{eR}]_{22})}{m_{\mathcal{A}}^2}-i\fr{[\Ga^{H}_{dR}]_{23}\mathrm{Im}([\Ga^{H}_{eR}]_{22})}{m_{H}^2}-i\fr{[\Ga^{H}_{dR}]_{23}\mathrm{Im}([\Ga^{\mathcal{H}}_{eR}]_{22})}{m_{\mathcal{H}}^2}\right], \\
C'_{P}&=&\fr{16\pi^2}{e^2}\fr{\sqrt{2}}{4G_FV^*_{ts}V_{tb}}\left[\fr{[\Ga^{\mathcal{A}}_{dR}]^*_{32}\mathrm{Re}([\Ga^{\mathcal{A}}_{eR}]_{22})}{m_{\mathcal{A}}^2}+i\fr{[\Ga^{H}_{dR}]^*_{32}\mathrm{Im}([\Ga^{H}_{eR}]_{22})}{m_{H}^2}+i\fr{[\Ga^{H}_{dR}]^*_{32}\mathrm{Im}([\Ga^{\mathcal{H}}_{eR}]_{22})}{m_{\mathcal{H}}^2}\right]. \label{WCBsmm}
\eea 

The branching ratio for the rare decay $B_s \to \mu^+\mu^-$ is given by
\bea 
\ep_{B_s\to \mu^+\mu^-}&=&\fr{\mathrm{BR}(B_s\to \mu^+\mu^-)_{\text{SM+NP}}}{\mathrm{BR}(B_s\to \mu^+\mu^-)_{\text{SM}}}=\fr{\mathrm{BR}(B_s\to \mu^+\mu^-)_{\text{exp}}}{\mathrm{BR}(B_s\to \mu^+\mu^-)_{\text{SM}}}\crn 
&=&\fr{1+\mathcal{A}_{\Delta \Ga}\,y_s}{1-y^2_s}\fr{|P|^2+|S|^2}{|C^\mathrm{SM}_{10}|^2},\eea
where the parameters $P$ and $S$ are defined as 
\bea 
P&=&C^{\text{SM}}_{10}-C'_{10}+\fr{m_{B_s}^2}{2m_{\mu}(m_b+m_s)}(C_P-C'_P), \\  S&=&\sqrt{1-\fr{4m_{\mu}^2}{m_{B_s}^2}}\fr{m_{B_s}^2}{2m_{\mu}(m_b+m_s)}(C_S-C'_S).
\eea 
Here, the observable $\mathcal{A}_{\Delta \Ga}$ takes the value $1$ in the SM. In the presence of NP contributions, it is modified as follows~\cite{Buras:2013uqa}
\be  
\mathcal{A}_{\Delta \Ga}=\fr{|P|^2\cos{(2\phi_P-\phi_s^{\text{NP}})}-|S|^2\cos{(2\phi_S-\phi_s^{\text{NP}})}}{|P|^2+|S|^2}.
\ee 
To estimate the impact of NP, we consider the predicted ratio $\ep_{B_s\to \mu^+\mu^-}$ constrained by the experimental measurement within the corresponding $2\sigma$ range, 
\be 
\fr{\mathrm{BR}(B_s\to \mu^+\mu^-)_{\text{exp}}}{\mathrm{BR}(B_s\to \mu^+\mu^-)_{\text{SM}}}\in [0.7574,1.0778], \label{Bsmm_const}
\ee 
where the experimental branching ratio is $\mathrm{BR}(B_s\to \mu^+\mu^-)_{\text{exp}}=3.34(27)\times 10^{-9}$~\cite{Czaja:2024the}, and the SM prediction, including power-enhanced QED corrections, is $\mathrm{BR}(B_s\to \mu^+\mu^-)_{\text{SM}}=3.64(12)\times 10^{-9}$~\cite{ParticleDataGroup:2024cfk}.

\begin{figure}[t]
	\includegraphics[scale=0.3]{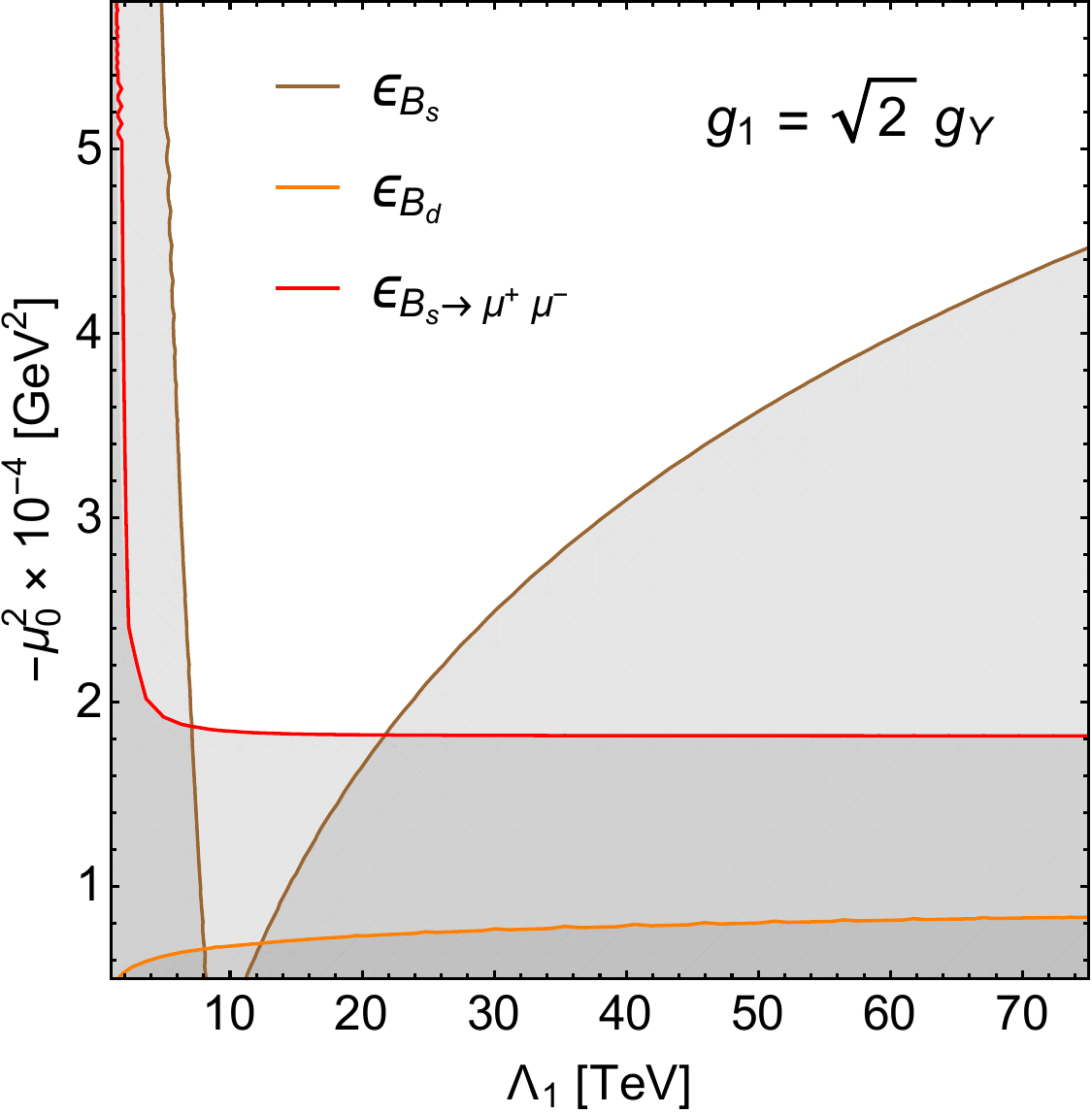}
	\includegraphics[scale=0.3]{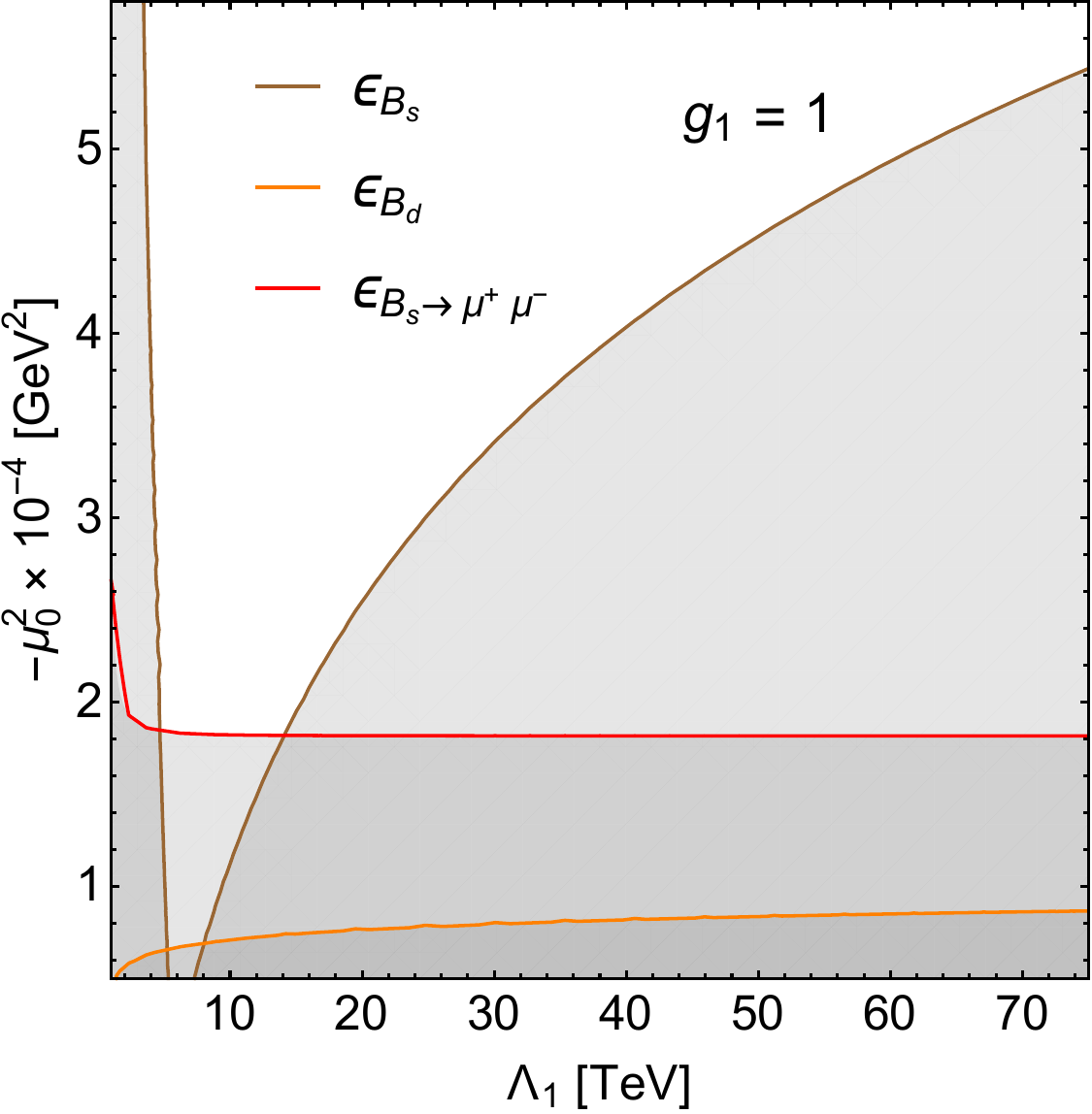}
	\includegraphics[scale=0.3]{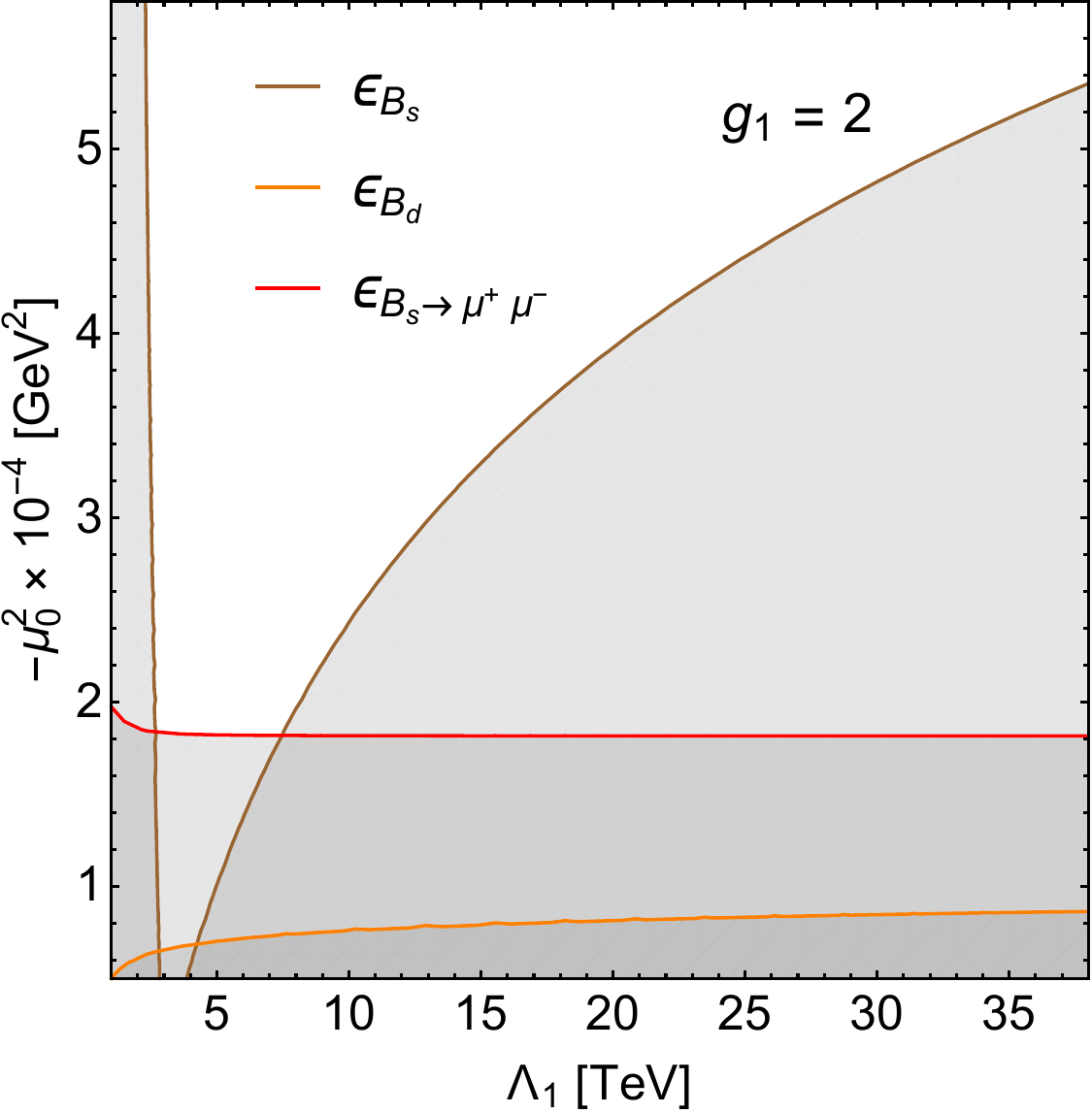}
	\caption{Lower bounds of $-\mu_{0}^2$ as a function of the new-physics scale $\La_1$ for several values of $g_1$. The gray-shaded region is excluded by the $2\sigma$ flavor constraints, while the white region is allowed.}
	\label{FCNC}
\end{figure}

For the numerical analysis, we fix $z=1$, $\La_1=\La_2$, and $v_2=0.52$~GeV, consistent with the fit to the SM fermion mass spectrum and mixing parameters. Fig.~\ref{FCNC} illustrates the dependence of the lower bound of the Higgs coupling $-\mu_{0}^2$ on the new-physics scale $\La_1$ for several values of the $U(1)_\mathcal{Y}$ gauge coupling $g_1$.\footnote{The gauge couplings $g_1$ and $g_2$ are related by $g_2=g_1g_Y/\sqrt{g_1^2-g_Y^2}$, requiring $g_1\geq g_Y$ to ensure a real and positive value of $g_2$.} Here, the gray-shaded region is excluded by the $2\sigma$ experimental constraints from $\Delta m_{B_q}$ and $\mathrm{BR}(B_s\to \mu^+\mu^-)$, given in Eqs.~(\ref{Bq_const1}), (\ref{Bq_const2}), and (\ref{Bsmm_const}), while the white region remains allowed. For $\La_1$ in the few-TeV range, the lower bound $-\mu_{0}^2\gtrsim 1.84\times 10^4~\mathrm{GeV}^2$ is controlled by the constraint from $\ep_{B_s\to \mu^+\mu^-}$, which depends only mildly on $\La_1$ and $g_1$. This value is consistent with the theoretical expectation $-\mu_0^2\sim \mathcal{O}(10^4)~\mathrm{GeV}^2$, estimated from the loop correction to the $\mu_0^2\Ph_1^{\dagger}\Ph_2$ term in the scalar potential (see Appendix~\ref{poten}). At larger $\La_1$, however, the lower bound on $-\mu_0^2$ is instead governed by the $\ep_{B_s}$ constraint, which exhibits a stronger dependence on both $\La_1$ and $g_1$. Although the direct $Z'$-mediated contributions---through the Wilson coefficients $C^{q}_{RR}$ and $C'_{10}$---are numerically much smaller than the scalar-mediated ones ($H,\mathcal{H}, \mathcal{A}$), the overall $B_q$--$\bar B_q$ mixing amplitude still retains a nontrivial dependence on $\La_1$ through the factor $\eta_{B_q}$. As $\La_1$ increases, this dependence requires a larger magnitude of $-\mu_0^2$ to compensate and keep the total mixing amplitude within the experimentally allowed range. Finally, Fig.~\ref{FCNC} also implies a conservative lower bound on the new-physics scale, $\La_1\gtrsim \mathcal{O}(1)~\mathrm{TeV}$.
 
\subsection{Collider signatures}
\subsubsection{$Z^{\prime }$ Production at Proton--Proton Colliders}
We analyze the Drell--Yan production of the heavy $Z^{\prime }$ gauge boson
at the LHC. The $Z^{\prime }$ gauge boson is primarily produced at the LHC
through the Drell--Yan process, with the light quarks $u$, $d$, $s$ providing
the dominant  parton distribution functions (PDF) contributions. The total $pp\rightarrow Z^{\prime }$ cross
section at $\sqrt{S}$ energy, mediated by $q\bar{q}$ annihilation can be
expressed as: 
\begin{eqnarray}
\sigma _{pp\rightarrow Z^{\prime }}^{\left(\mathrm{DrellYan}\right) }(S) &=&\frac{%
g^{2}\pi }{12c_{W}^{2}S}\left\{ \left[ \left( g_{V}^{Z_{2}}\left( u_{\alpha
}\right) \right) ^{2}+\left( g_{A}^{Z_{2}}\left( u_{\alpha }\right) \right)
^{2}\right] \int_{\ln \sqrt{\frac{m_{Z^{\prime }}^{2}}{S}}}^{-\ln \sqrt{%
\frac{m_{Z^{\prime }}^{2}}{S}}}f_{p/u}\left( \sqrt{\frac{m_{Z^{\prime }}^{2}%
}{S}}e^{y},\mu ^{2}\right) f_{p/\overline{u}}\left( \sqrt{\frac{m_{Z^{\prime
}}^{2}}{S}}e^{-y},\mu ^{2}\right) dy\right.   \notag \\
&&+\left. \left[ \left( g_{V}^{Z_{2}}\left( d_{\alpha }\right) \right)
^{2}+\left( g_{A}^{Z_{2}}\left( d_{\alpha }\right) \right) ^{2}\right]
\int_{\ln \sqrt{\frac{m_{Z^{\prime }}^{2}}{S}}}^{-\ln \sqrt{\frac{%
m_{Z^{\prime }}^{2}}{S}}}f_{p/d}\left( \sqrt{\frac{m_{Z^{\prime }}^{2}}{S}}%
e^{y},\mu ^{2}\right) f_{p/\overline{d}}\left( \sqrt{\frac{m_{Z^{\prime
}}^{2}}{S}}e^{-y},\mu ^{2}\right) dy\right.   \notag \\
&&+\left. \left[ \left( g_{V}^{Z_{2}}\left( d_{\alpha }\right) \right)
^{2}+\left( g_{A}^{Z_{2}}\left( d_{\alpha }\right) \right) ^{2}\right]
\int_{\ln \sqrt{\frac{m_{Z^{\prime }}^{2}}{S}}}^{-\ln \sqrt{\frac{%
m_{Z^{\prime }}^{2}}{S}}}f_{p/s}\left( \sqrt{\frac{m_{Z^{\prime }}^{2}}{S}}%
e^{y},\mu ^{2}\right) f_{p/\overline{s}}\left( \sqrt{\frac{m_{Z^{\prime
}}^{2}}{S}}e^{-y},\mu ^{2}\right) dy\right\},
\end{eqnarray}
where the functions $f_{p/u}\left( x_1,\mu^2 \right)$ ($f_{p/\overline{u}}\left(x_2,\mu ^2 \right)$), $f_{p/d}\left(x_1,\mu
^2 \right)$ ($f_{p/\overline{d}}\left( x_2,\mu^2 \right)$) and $f_{p/s}\left(x_1,\mu
^2 \right)$ ($f_{p/\overline{s}}\left( x_2,\mu^2 \right)$) correspond to the parton distribution functions of the light up, down and strange quarks (antiquarks), respectively, in the proton which carry momentum fractions $x_1$ ($x_2$) of the proton. The factorization scale is taken to be $\mu =m_{Z^\prime }$.

\begin{figure}[t]
	\includegraphics[width=7cm, height=5cm]{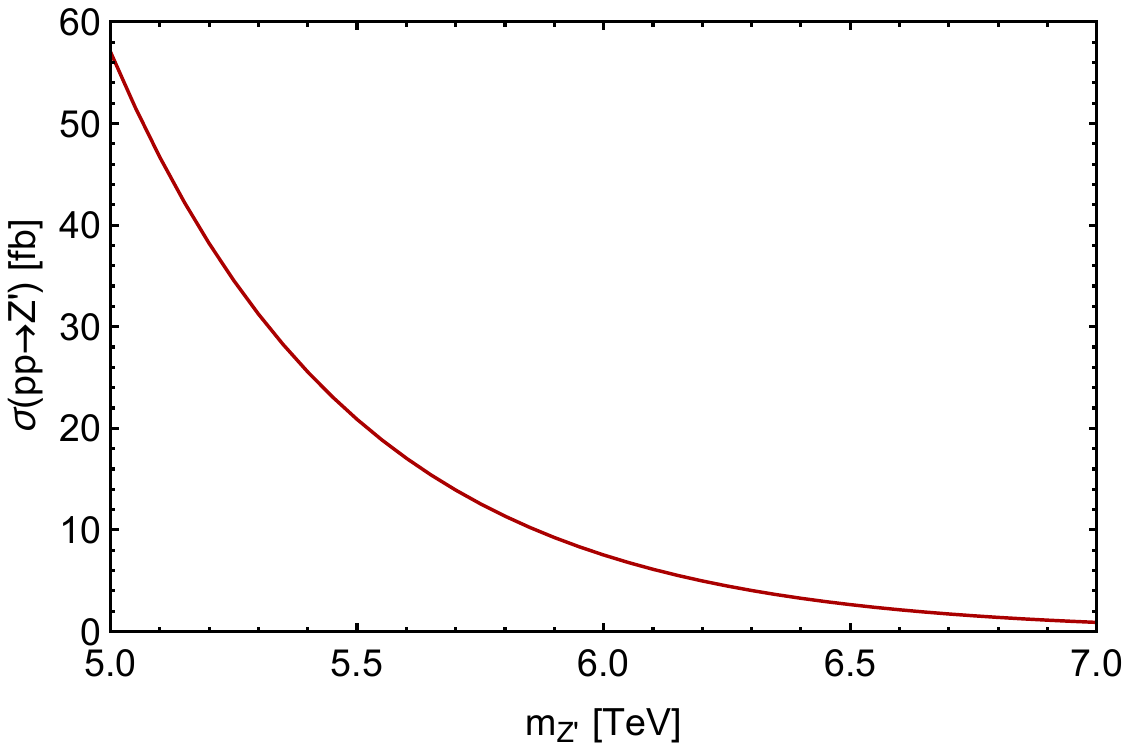}\includegraphics[width=7cm, height=4.85cm]{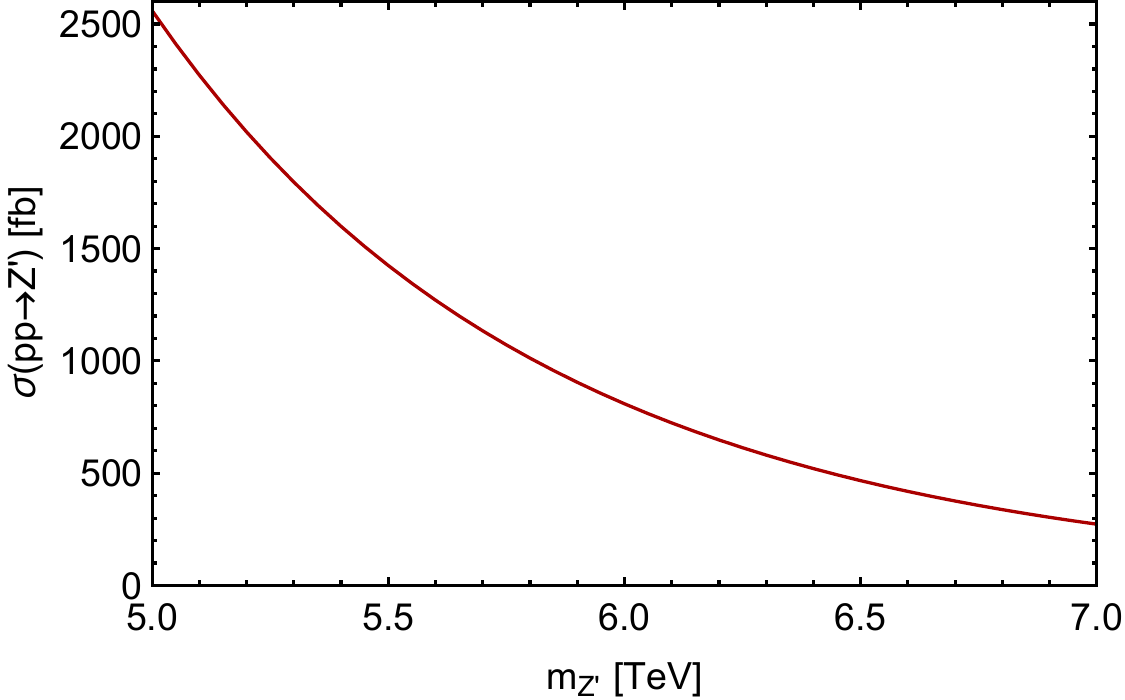}
\caption{Total cross section for $Z^\prime $ production via Drell--Yan mechanism at the LHC as a function of the $Z^\prime $ mass, for $\protect\sqrt{S}=14$ TeV (left) and $\protect\sqrt{S}=28$ TeV (right).}
	\label{qqtoZprime}
	\end{figure}
\begin{figure}[t]
	\includegraphics[width=7cm, height=5cm]{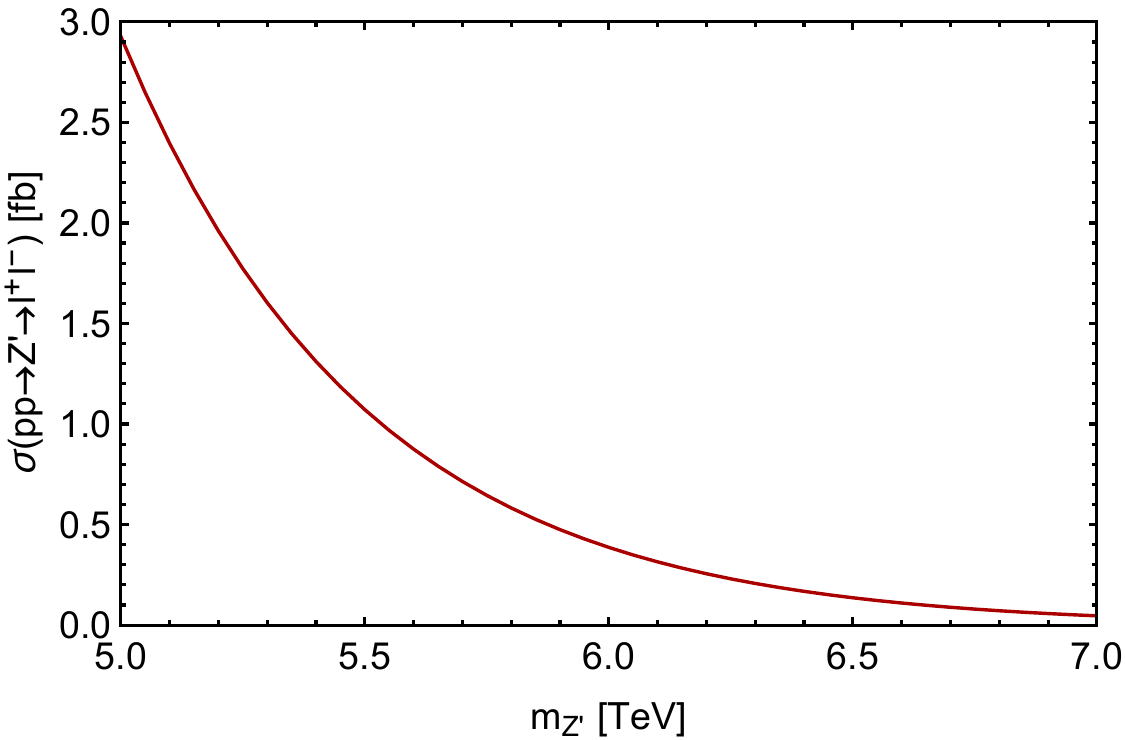}\includegraphics[width=7cm, height=5cm]{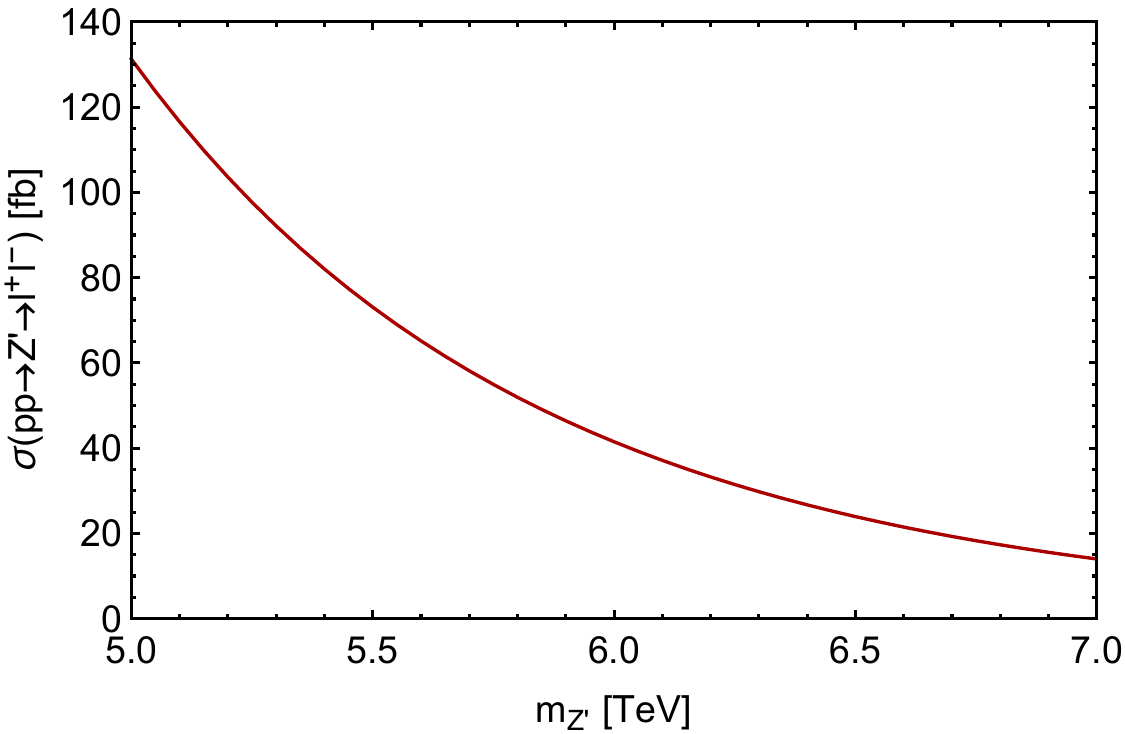}
	\caption{Total cross section for resonant $pp\rightarrow Z^\prime\rightarrow l^{+}l^{-}$ production via Drell--Yan mechanism at the LHC as a function of the $Z^\prime $ mass, for $\protect\sqrt{S}=14$ TeV (left) and $\protect\sqrt{S}=28$ TeV (right).}
	\label{qqtoZprimetoll}
\end{figure}

Fig.~\ref{qqtoZprime} displays the total cross section at the LHC for the $Z^\prime $ production via the Drell--Yan mechanism for $\protect\sqrt{S}=14$ TeV (left plot) and $\protect\sqrt{S}=28$ TeV (right plot) as a function of the $Z^\prime $ mass $m_{Z'}$, which is taken to range from $5$ TeV up to $7$ TeV. Here we have set $z=1$, $\theta=0.45\pi$ and $\Lambda_1=\Lambda_2=10$ TeV. We focus on values of $Z^{\prime }$ gauge boson mass above $5$ TeV, choice compatible with the constraint $\frac{M_{Z^{\prime }}}{g}>7$ TeV ($g\approx 0.65$), derived from LEP I/II measurements of $e^{+}e^{-}\rightarrow l^{+}l^{-}$~\cite{LEP:2004xhf,Carena:2004xs,Das:2021esm} and supported by LHC searches~\cite{ATLAS:2019erb,CMS:2021ctt}. Further bounds on $\frac{M_{Z^{\prime }}}{g}$ for LEP II and projected scenarios at future $e^{+}e^{-}$ colliders (e.g., ILC) are analyzed in Ref.~\cite{Das:2021esm}. For our analysis, we apply the conservative LEP II limit, as other bounds depend on unrealized future experiments. Within the above mentioned $Z^\prime$ mass range, the total production cross section spans $57$ fb to $0.9$ fb at $\sqrt{S}=14$ TeV. For the $28$ TeV LHC upgrade, the cross section rises from $2560$ fb down to $273$ fb as seen in the right panel of Fig.~\ref{qqtoZprime}. The $Z^\prime$ predominantly decays into SM fermion pairs, making dilepton or dijet excesses a potential discovery channel at the LHC. The total cross sections for the $pp\rightarrow Z^\prime\rightarrow l^{+}l^{-}$ resonance production via Drell--Yan mechanism at a proton--proton collider for $\protect\sqrt{S}=14$ TeV and $\protect\sqrt{S}=28$ TeV are shown in the left and right panels of Fig.~\ref{qqtoZprimetoll}, respectively. The corresponding cross sections vary from $2.9$~fb to $0.05$~fb at $\sqrt{S}=14$~TeV, and from $131$~fb to $14$~fb at $\sqrt{S}=28$~TeV.

\subsubsection{$\mathcal{H}$ Scalar Production at Proton--Proton Colliders}
\label{sec:HeavyScalar}

\begin{figure}[h]
\includegraphics[width=7cm, height=5cm]{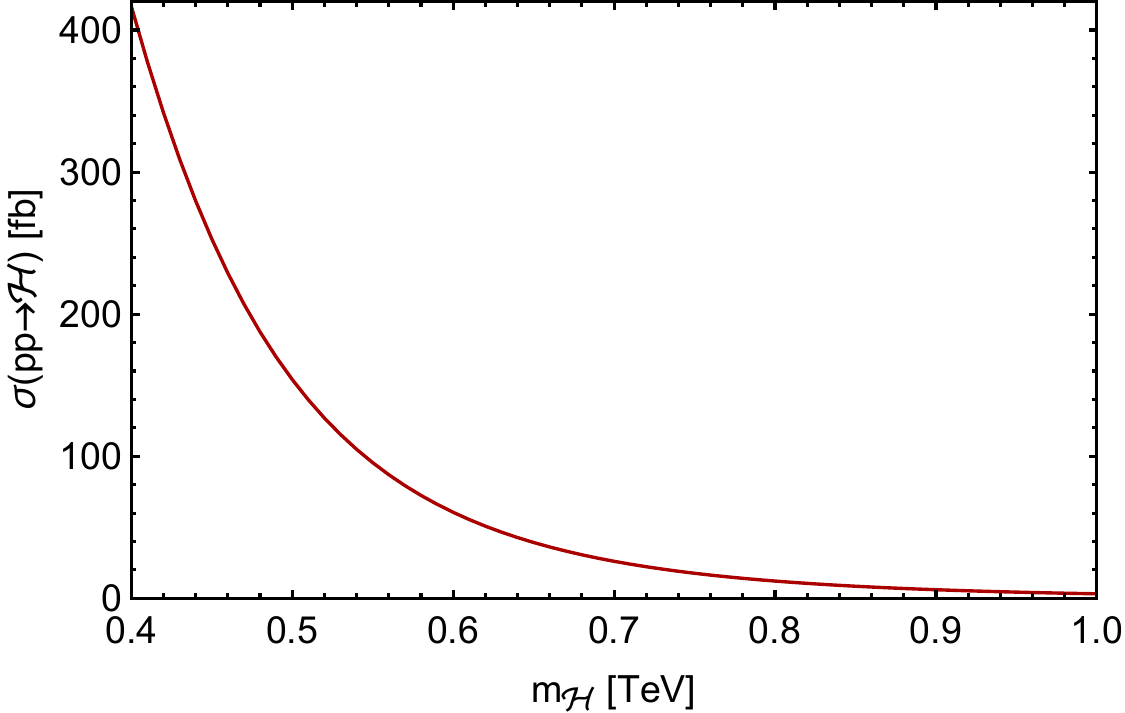}\includegraphics[width=7cm, height=5cm]{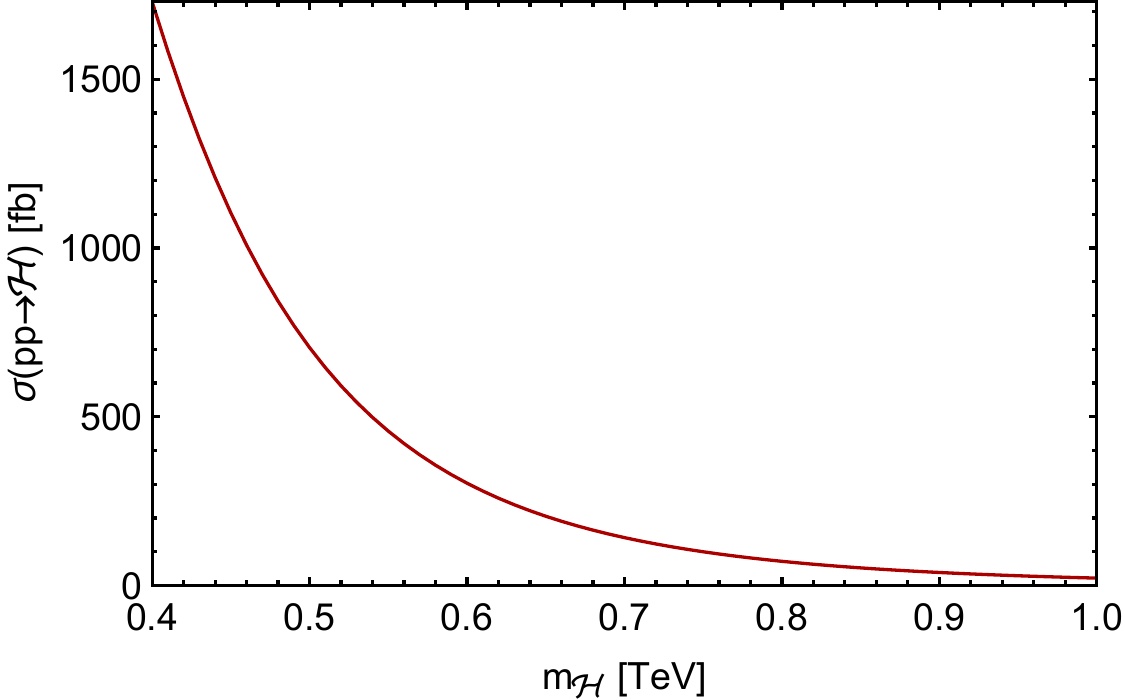}
		\caption{Total cross section for $\mathcal{H}$ production via gluon fusion at a proposed $pp$ collider as a function of $m_{\mathcal{H}}$, for $\protect\sqrt{S}=14$ TeV (left) and $\protect\sqrt{S}=28$ TeV (right).}
		\label{ggtoheavyH}
	\end{figure}

We now examine the production mechanisms for the heavy neutral scalar, denoted $\mathcal{H}$, in proton--proton ($pp$) collisions. The gluon fusion process represents the most significant production channel at the LHC, proceeding through a top and charm quark loops. The corresponding cross section for $pp \to \mathcal{H}$ at a center-of-mass energy $\sqrt{S}$ is expressed as:
\begin{eqnarray}
\sigma _{pp\rightarrow gg\rightarrow \mathcal{H}}\left( S\right) &=&\frac{\alpha
	_{S}^{2}m_{\mathcal{H}}^{2}}{64\pi S}\left[\frac{a_{\mathcal{H}t\bar{t}}}{v_1}I\left( \frac{m_{\mathcal{H}}^{2}}{m_{t}^{2}}%
\right)+\frac{a_{\mathcal{H}c\bar{c}}}{v_2}I\left( \frac{m_{\mathcal{H}}^{2}}{m_{c}^{2}}%
\right)\right]^{2}\notag\\
&&\times\int_{\ln \sqrt{\frac{m_{\mathcal{H}}^{2}}{S}}}^{-\ln \sqrt{\frac{%
			m_{\mathcal{H}}^{2}}{S}}}f_{p/g}\left( \sqrt{\frac{m_{\mathcal{H}}^{2}}{S}}e^{y},\mu
^{2}\right)f_{p/g}\left( \sqrt{\frac{m_{\mathcal{H}}^{2}}{S}}e^{-y},\mu^{2}\right) dy,
\end{eqnarray}
where the functions $f_{p/g}(x_1, \mu^2)$ and $f_{p/g}(x_2, \mu^2)$ describe the gluon density within the proton, carrying momentum fractions $x_1$ and $x_2$, respectively. The factorization scale is chosen as $\mu = m_{\mathcal{H}}$. The loop function $I(z)$, which arises from the top-quark loop, is given by the integral:
\begin{equation}
I(z)=\int_{0}^{1}dx\int_{0}^{1-x}dy\frac{1-4xy}{1-zxy}.
\label{g1a}
\end{equation}
The resulting production cross sections are displayed in the subsequent figures. The dependence of the $\mathcal{H}$ production cross section on its mass is plotted in the left panel of Fig.~\ref{ggtoheavyH} for the LHC operating at $\sqrt{S} = 14$ TeV, considering a mass range from $0.4$ TeV to $1$ TeV. For this calculation, the Yukawa couplings $a_{\mathcal{H}t\bar{t}}$ and $a_{\mathcal{H}c\bar{c}}$ are fixed to $0.1$ and $1$, respectively. Besides that we have set $v_2=0.5$ GeV and $v_1=\sqrt{v^2-v^2_2}\approx 246$ GeV. Within this mass window, the cross section falls from about $417$ fb to $3.2$ fb.

Significant enhancements in the production rate are anticipated at higher collision energies. The right panel of Fig.~\ref{ggtoheavyH} illustrates the cross section for a potential future LHC run at $\sqrt{S} = 28$ TeV, where it spans from $1730$ fb down to $22$ fb across the same $\mathcal{H}$ mass range. From a phenomenological perspective, if $\mathcal{H}$ is the lightest beyond-the-Standard-Model scalar, it decays predominantly to pairs of first- and second-generation quarks or charged leptons. These decay channels would yield final states rich in jets and/or leptons. It is crucial to note, however, that our analysis is conducted near the alignment limit, where the coupling of the SM light Higgs boson $H$ are very close to the SM expectation. Consequently, the contribution from $\mathcal{H}$ pair production to multilepton and multijet signals is predicted to be minor compared to the dominant SM processes.

\section{\label{darkmatter}Complex scalar singlet dark matter}
This model features a residual parity $P_\mathsf{D}$ under which the scalars $\phi^0$, $\eta_1$, and $\eta_2$ are odd. The conserved residual parity $P_\mathsf{D}$ stabilizes the lightest such state, identifying it as a DM candidate. The mass of this candidate is naturally heavy (typically TeV-scale), being determined by the breaking scales $\La_{1,2}$. The DM interacts with SM particles through contact terms in the scalar potential, as well as via scalar and gauge portals. Consequently, its relic abundance is governed by the thermal freeze-out mechanism, implying WIMP-like behavior. 
For simplicity, we assume the DM candidate to be the lightest state in the spectrum beyond the SM, with the possible exception of the physical Goldstone boson $\mathcal{G}$.

\subsection{Dark matter as a complex scalar singlet $\eta_2$}
We first consider the scenario in which the complex scalar singlet $\eta_2$ acts as the DM candidate. Since $\eta_2$ is an $SU(2)_L$ singlet with both $\mathcal{Y}=0$ and $\mathcal{R}=0$, the $Z_1$ and $Z_2$ annihilation portals are absent. The $Z_\mathsf{R}$ portal is also negligible due to the mass hierarchy, namely $m_{\eta_2}\ll m_{Z_\mathsf{R}}$. Furthermore, the $\mathcal{H}_{1,2}$ portals can be ignored under the assumption $|\la_{30,31}|\ll 1$ as they do not contribute to direct DM detection. Consequently, $\eta_2$ predominantly annihilates into SM particles through the contact interaction $\fr 1 2 \la_{28}\eta^*_2\eta_2H^2$, as well as via the scalar portals mediated by $H$ and $\mathcal{H}$. The corresponding thermally averaged annihilation cross section is then given by:
\bea \langle\sigma v_\text{rel}\rangle_{\eta^*_2\eta_2} &=& \langle\sigma v_\text{rel}\rangle_{\eta^*_2\eta_2\to HH} + \langle\sigma v_\text{rel}\rangle_{\eta^*_2\eta_2\to ff^c}+ \langle\sigma v_\text{rel}\rangle_{\eta^*_2\eta_2\to W^+W^-,Z_1Z_1}\crn
&\simeq&\frac{1}{32\pi m_{\eta_2}^2}\left[\la_{28}+\frac{\la^2\La^2_2}{4(m^2_{\eta_2}+m^2_{\phi^0})}+\frac{3\la_{28}\la_1v_1^2}{4m^2_{\eta_2}}+\frac{\la_{29}(\la_3+\la_4)v_2^2}{4m^2_{\eta_2}-m^2_\mathcal{H}}\right]^2\crn
&&+\frac{3m_t^2}{\pi}\left[\frac{\la_{28}}{4m_{\eta_2}^2}+\frac{\la_{29}\ep_1 v_2}{(4m_{\eta_2}^2-m_\mathcal{H}^2)v_1}\right]^2+\frac{3}{16\pi m^2_{\eta_2}v^4}\left(\la_{28}v_1^2+\frac{4\la_{29}v_2^2m_{\eta_2}^2}{4m_{\eta_2}^2-m_\mathcal{H}^2}\right)^2.  \eea

The scattering of $\eta_2$ off a nuclear target is primarily mediated by the $t$-channel exchange of the scalar bosons $H$ and $\mathcal{H}$. The effective Lagrangian describing this interaction at the quark level is given by
\be\mathcal{L}^\text{eff,scalar}_{\eta_2\text{--quark}}= C^S_{\eta_2q}\eta^*_2\eta_2\bar{q}q=m_q\left(\frac{C_q^H}{m^2_H}+\frac{C_q^\mathcal{H}}{m^2_\mathcal{H}}\right)\eta^*_2\eta_2\bar{q}q, \ee
where $m_q$ is the mass of the quark $q$, and 
\bea C_u^H &=& C_c^H=C_d^H= C_s^H \simeq \frac{\la_{28}\ep_1v_1}{v_2},\hs C_t^H=C_b^H\simeq -\la_{28},\\
C_u^\mathcal{H} &=& C_c^\mathcal{H}=C_d^\mathcal{H}= C_s^\mathcal{H}= -\fr{\la_{28}\ep_1v_1+\la_{29}v_2}{v_2},\hs C_t^\mathcal{H}=C_b^\mathcal{H}= -\frac{\ep_1(\la_{28}\ep_1v_1+\la_{29}v_2)}{v_1}. \eea
Then, at the nucleon level, the effective Lagrangian takes the form
\be\mathcal{L}^\text{eff,scalar}_{\eta_2\text{--nucleon}}= C_{\eta_2N}\eta^*_2\eta_2\bar{N}N, \ee
in which $C_{\eta_2N}~(N=p,n)$ denotes the effective coupling of DM to nucleons (protons, neutrons). These effective couplings are given by
\be C_{\eta_2N} = \frac{m_N}{2m_{\eta_2}}\left[\sum_{q=u,d,s}f^N_{Tq}\frac{C^S_{\eta_2q}}{m_q}+\fr{2}{27}\left(1-\sum_{q=u,d,s}f^N_{Tq}\right)\sum_{q=c,b,t}\frac{C^S_{\eta_2q}}{m_q}\right],\ee 
where $f^N_{Tq}$ are the scalar form factors of the nucleon for the quark $q$. The number values used in our analysis are $f^{p(n)}_{Tu}\simeq 0.0208(0.0189)$, $f^{p(n)}_{Td}\simeq 0.0411(0.0451)$, and $f^{p(n)}_{Ts}\simeq 0.043(0.043)$~\cite{Junnarkar:2013ac,Hoferichter:2015dsa}.
From this effective interaction, the spin-independent (SI) scattering cross-section of $\eta_2$ off a nuclear target $\mathbb{A}$ is given by~\cite{Jungman:1995df}
\be \sigma^\text{SI}_{\eta_2\text{--}\mathbb{A}}=\frac{\mu_\mathbb{A}^2}{\pi}[ZC_{\eta_2p}+(A-Z)C_{\eta_2n}]^2,\ee
where $\mu_\mathbb{A}=\fr{m_\mathbb{A}m_{\eta_2}}{m_\mathbb{A}+m_{\eta_2}}$ is the reduced mass of the DM--nucleus system. Here, $m_\mathbb{A}$ denotes the mass of the target nucleus, $Z$ is the atomic number, and $A$ is the mass number.

We find that the ratio $C_{\eta_2n}/C_{\eta_2p}$ lies within the range of approximately $0.966$--$1.018$ for representative parameter values: $\la_{29}=0.1$, $v_2=0.52$~GeV, $m_{\eta_2}\sim\mathcal{O}(1)$~TeV, $\la_{28}=0.001$--1, and $-\mu_0^2=(0.4\text{--}10)\times 10^4~\mathrm{GeV}^2$. Accordingly, the SI DM-nucleon cross section, $\sigma^\text{SI}_{\eta_2}$, is related to the DM--nucleus cross section, $\sigma^\text{SI}_{\eta_2\text{--}\mathbb{A}}$, by $\sigma^\text{SI}_{\eta_2}=(\sigma^\text{SI}_{\eta_2\text{--}\mathbb{A}}/A^2)(\mu_p/\mu_\mathbb{A})^2$, where $\mu_p$ is the reduced mass of the DM--proton system~\cite{Feng:2013vod}. 

\begin{figure}[h]
    \centering
    \includegraphics[width=0.3\textwidth]{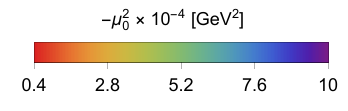}\\
    \vspace{0em} 
     \includegraphics[width=8.5cm, height=6cm]{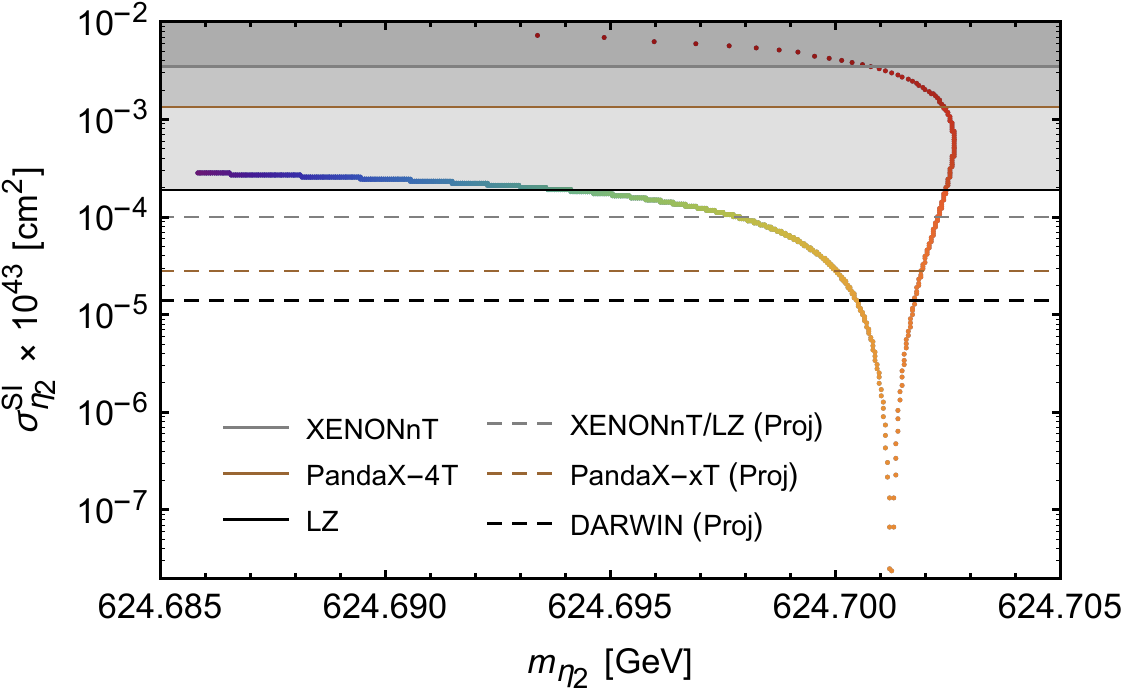}
     \includegraphics[width=8.5cm, height=6cm]{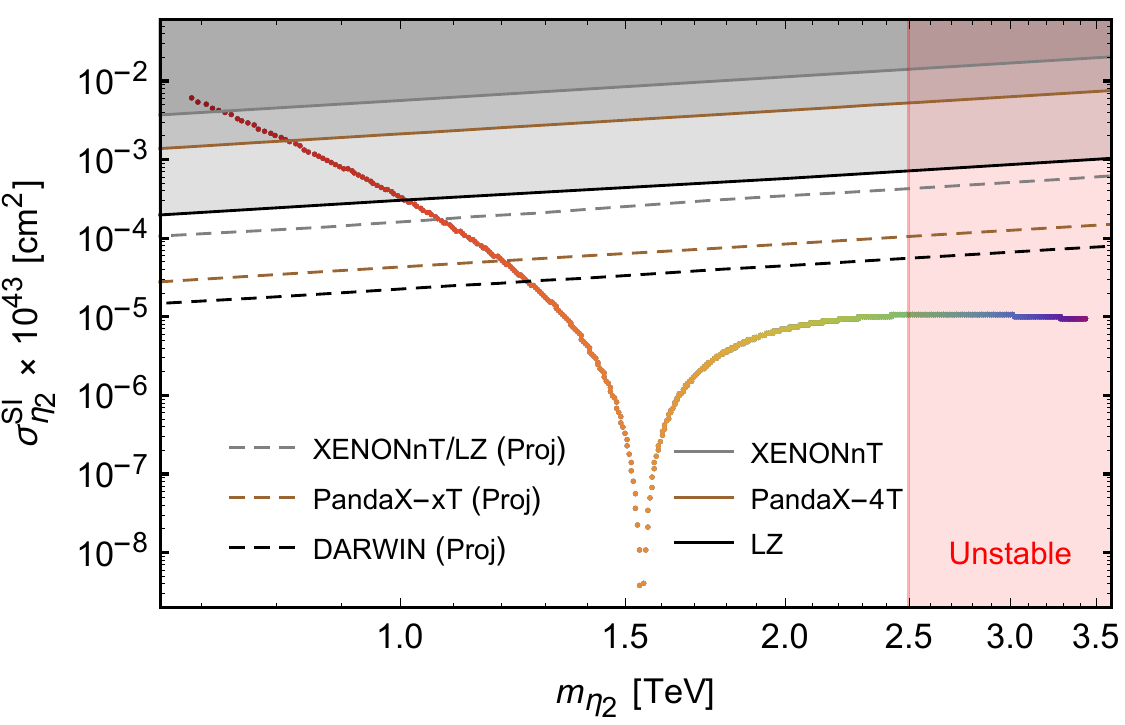}
     \caption{The SI cross section of the DM candidate $\eta_2$ with nucleons as a function of its mass $m_{\eta_2}$, shown for various values of $\mu_0^2$. The shaded black regions are excluded by the XENONnT~\cite{XENON:2023cxc}, PandaX-4T~\cite{PandaX:2024qfu}, and LZ~\cite{LZ:2024zvo} experiments, while the shaded red region is excluded due to the instability of $\eta_2$. Projected sensitivities from future direct-detection experiments---XENONnT/LZ~\cite{XENON:2020kmp,LZ:2018qzl}, PandaX-xT~\cite{PANDA-X:2024dlo}, and DARWIN~\cite{DARWIN:2016hyl}---are indicated by the green, brown, and red dot-dashed curves, respectively.}
\label{DMeta2}
\end{figure}

Fig.~\ref{DMeta2} shows the correlation between the SI cross section $\sigma^\text{SI}_{\eta_2}$ and the DM mass $m_{\eta_2}$, yielding the correct DM relic abundance, $\Om_{\eta_2}h^2\simeq 0.1~\mathrm{pb}/\langle\sigma v_\text{rel}\rangle_{\eta^*_2\eta_2}\simeq 0.12$, as measured by the Planck experiment~\cite{Planck:2018vyg}. The numerical results are obtained by setting $\la_{28}=\la_{29}=\la_3=\la_4=\la=0.1$, $v_2=0.52$~GeV, $\La_2=10$~TeV, $m_{\phi^0}=2.5$~TeV, $Z=54$, and $A=131$, with $-\mu_0^2$ varying in the range $(0.4\text{--}10)\times 10^4~\mathrm{GeV}^2$. In the left panel, the mass region consistent with the relic density corresponds to the $H$--portal, which is nearly independent of $\mu_0^2$, leading to a relatively narrow allowed region. By contrast, the right panel shows the mass region associated with the $\mathcal{H}$--portal, which is strongly sensitive to $\mu_0^2$ and results in a much broader allowed parameter space. For comparison, the most stringent experimental bounds from XENONnT~\cite{XENON:2023cxc}, PandaX-4T~\cite{PandaX:2024qfu}, and LZ~\cite{LZ:2024zvo} are also displayed. The shaded black regions are excluded by these experiments, while the red-shaded region is excluded due to DM instability. Two viable mass regions for $\eta_2$ are identified: $m_{\eta_2}\simeq 624.7$~GeV for $-\mu_0^2\simeq (1.23\text{--}5.86)\times 10^4~\mathrm{GeV}^2$ (left panel), and $m_{\eta_2}\simeq (1.01$--$2.50)$~TeV for $-\mu_0^2\simeq (0.87\text{--}5.28)\times 10^4~\mathrm{GeV}^2$ (right panel). Furthermore, the projected sensitivities of upcoming direct-detection experiments imply even tighter constraints: $-\mu_0^2\simeq (1.72\text{--}2.46)\times 10^4~\mathrm{GeV}^2$ (left panel), and $m_{\eta_2}\simeq (1.26$--$2.50)$~TeV for $-\mu_0^2\simeq (1.35\text{--}5.28)\times 10^4~\mathrm{GeV}^2$ (right panel)~\cite{XENON:2020kmp,LZ:2018qzl,PANDA-X:2024dlo,DARWIN:2016hyl}. Finally, the smallest value of $\sigma_{\eta_2}^\text{SI}$ is found at $-\mu_0^2\simeq 2.02\times 10^{4}~\mathrm{GeV}^2$ (both panels), and $m_{\eta_2}\simeq 1.55~\mathrm{TeV}$ (right panel). 

\subsection{Dark matter as a complex scalar singlet $\eta_1$}
We now consider the scenario where the complex scalar singlet $\eta_1$ plays the role of the DM particle. Compared to the $\eta_2$ candidate, $\eta_1$ has addition vector interactions because it carries $\mathcal{Y}=-\mathcal{R}=-z/3$ and couples through the $Z$--$Z'$ mixing. These new interactions contribute to both DM pair annihilation and SI DM-nucleon scattering. The thermally averaged annihilation cross section of $\eta_1$ can be written as 
\be\langle\sigma v_\text{rel}\rangle_{\eta^*_1\eta_1} = \langle\sigma v_\text{rel}\rangle_{\eta^*_1\eta_1}^\text{scalar} + \langle\sigma v_\text{rel}\rangle_{\eta^*_1\eta_1}^\text{vector},\ee
where the first term comes to scalar interactions similar to those of $\eta_2$, while the second term originates from vector interactions,
\be\langle\sigma v_\text{rel}\rangle_{\eta^*_1\eta_1}^\text{vector} = \langle\sigma v_\text{rel}\rangle_{\eta^*_1\eta_1\to ff^c}^\text{vector}+ \langle\sigma v_\text{rel}\rangle_{\eta^*_1\eta_1\to W^+W^-,Z_1Z_1}^\text{vector}+ \langle\sigma v_\text{rel}\rangle_{\eta^*_1\eta_1\to HZ_1}^\text{vector}.\ee
It is worth noting that the annihilation cross sections of $\eta_1$ into $ff^c, W^+W^-$, and $HZ_1$, mediated by gauge bosons $Z_{1,2}$, are velocity suppressed and therefore not valid near s-channel resonances~\cite{Griest:1990kh}. Additionally, the annihilation cross section of $\eta_1$ into $Z_1Z_1$ can be approximately expressed as~\cite{Arcadi:2017kky}
\be \langle\sigma v_\text{rel}\rangle_{\eta^*_1\eta_1\to Z_1Z_1}^\text{vector}\simeq \frac{z^4s_\varepsilon^4(g_1^2+g_2^2)^2}{162\pi m_{\eta_1}^2},\ee
which is also strongly suppressed due to the smallness of $s_\varepsilon$.

For the SI DM--nucleon scattering, the effective Lagrangian describing the vector interactions at the quark level is written as
\be\mathcal{L}^\text{eff,vector}_{\eta_1\text{--quark}}= C^V_{\eta_1q}(\eta^*_1i\overlr_\mu\eta_1)\bar{q}\gamma^\mu q=-\frac{g}{2c_W}\fr{z\sqrt{g_1^2+g_2^2}}{3} \left(\frac{g_V^{Z_1}(q)s_\varepsilon}{m^2_{Z_1}}-\frac{g_V^{Z_2}(q)c_\varepsilon}{m^2_{Z_2}}\right)(\eta^*_1i\overlr_\mu\eta_1)\bar{q}\gamma^\mu q. \ee
At the nucleon level, the corresponding effective interaction takes the form
\be\mathcal{L}^\text{eff,vector}_{\eta_1\text{--nucleon}}= C^V_{\eta_1N}(\eta^*_1i\overlr_\mu\eta_1)\bar{N}\gamma^\mu N, \ee
where $C^V_{\eta_1N}$ denotes the vector-current effective couplings to nucleons. They are given by $C^V_{\eta_1p}=2C^V_{\eta_1u}+ C^V_{\eta_1d}$ and $C^V_{\eta_1n}=C^V_{\eta_1u}+ 2C^V_{\eta_1d}$.
Hence, the total effective couplings of $\eta_1$ and $\eta_1^*$ to nucleons can be written as
\be C_{\eta_1(\eta_1^*)p} = C^S_{\eta_1p}\pm C^V_{\eta_1p}, \hs C_{\eta_1(\eta_1^*)n}=C^S_{\eta_1n}\pm C^V_{\eta_1n},\ee
which differ between proton and neutron, as well as between $\eta_1$ and $\eta_1^*$. The scalar effective couplings $C^S_{\eta_1N}$ are similar those of $\eta_2$, i.e., $C^S_{\eta_1N}=C_{\eta_2N}|_{m_{\eta_2}\to m_{\eta_1},\la_{28}\to\la_{24},\la_{29}\to\la_{25}}$. The averaged SI cross section of $\eta_1$ and $\eta_1^*$ scattering off a nucleus $\mathbb{A}$ is then given by
\be \sigma^\text{SI}_{\eta_1\text{--}\mathbb{A}}=\frac{\mu_\mathbb{A}^2}{2\pi}\{[ZC_{\eta_1p}+(A-Z)C_{\eta_1n}]^2+[ZC_{\eta_1^*p}+(A-Z)C_{\eta_1^*n}]^2\}.\ee
Accordingly, the SI DM-nucleon scattering cross section is estimated as~\cite{Feng:2011vu,Lao:2020inc}
\be \sigma^\text{SI}_{\eta_1}= \frac{\mu_p^2C_{\eta_1p}^2}{2\pi}\frac{\sum_{i}\rho_i\mu^2_{\mathbb{A}_i}\{[Z+(A_i-Z)C_{\eta_1n}/C_{\eta_1p}]^2+[ZC_{\eta_1^*p}/C_{\eta_1p}+(A_i-Z)C_{\eta_1^*n}/C_{\eta_1p}]^2\}}{\sum_{i}\rho_i\mu^2_{\mathbb{A}_i}A^2_i},\ee
where the nuclear target consists off isotopes $A_i$ with fractional abundance $\rho_i$ (in percent). For example, the natural xenon isotopes have $A_i(\rho_i)=\{128(1.9),129(26),130(4.1),131(21),132(27)$, $134(10),136(8.9)\}$ for xenon isotopes~\cite{Feng:2011vu}.

\begin{figure}[h]
    \centering
    \includegraphics[width=0.3\textwidth]{plotlegend.pdf}\\
    \vspace{0em} 
     \includegraphics[width=8.5cm, height=6cm]{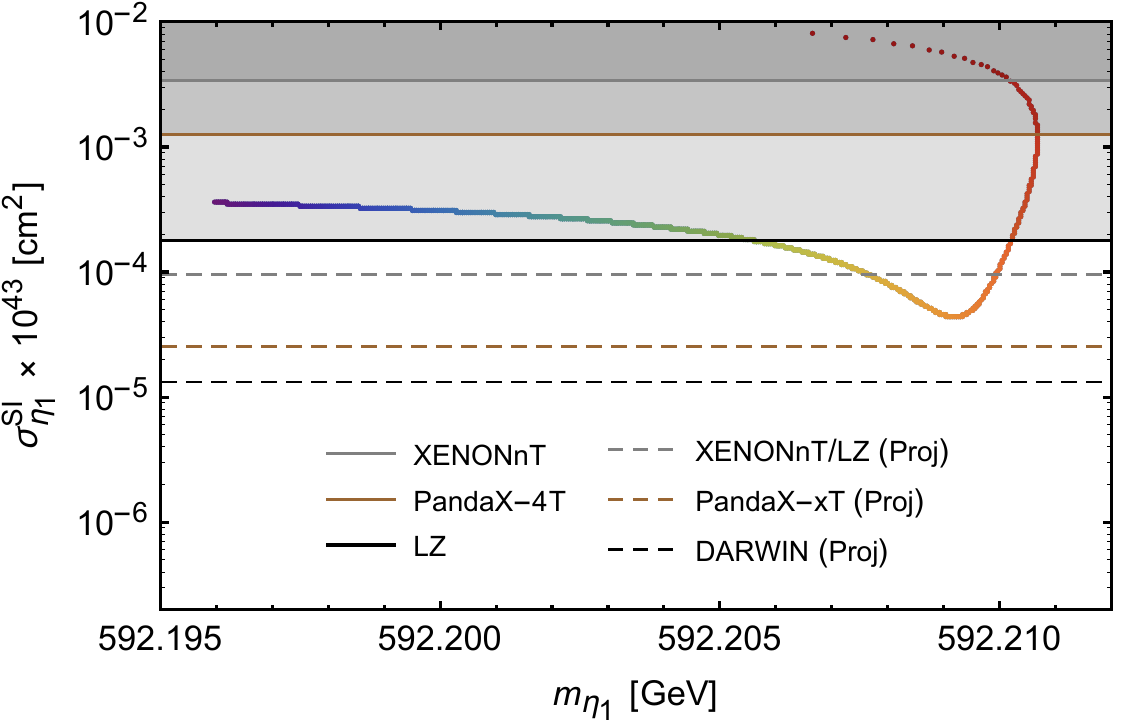}
     \includegraphics[width=8.5cm, height=6cm]{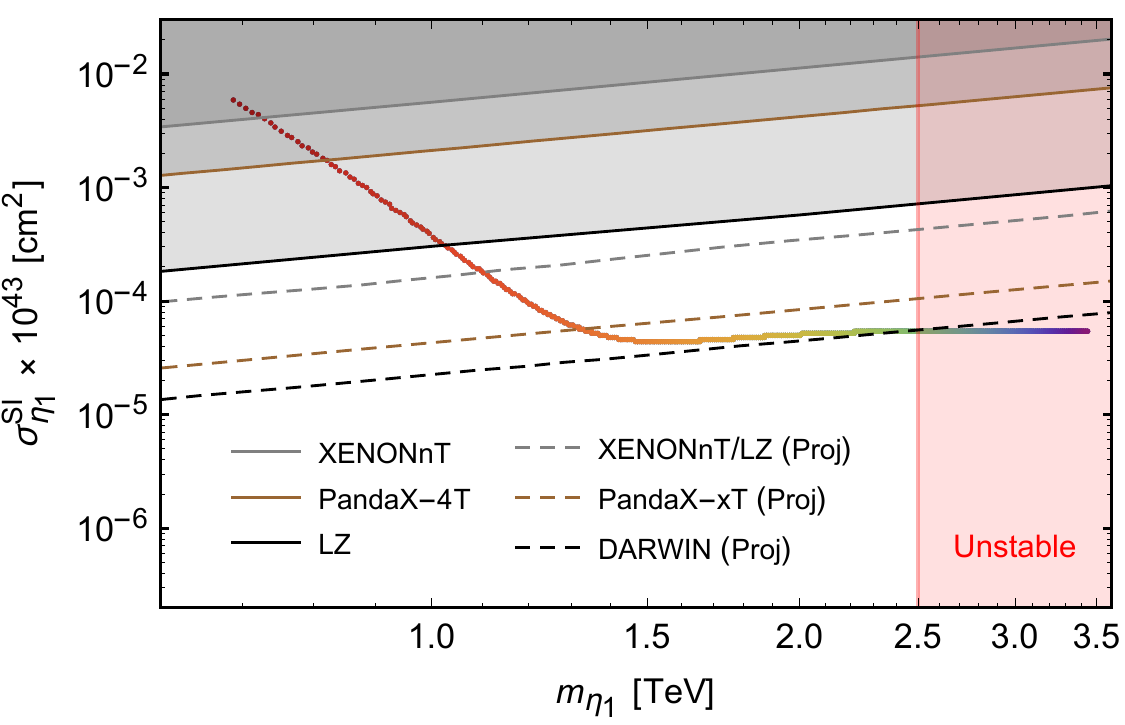}
     \caption{The same as Fig.~\ref{DMeta2}, except that the results correspond to the DM candidate $\eta_1$.}
\label{DMeta1}
\end{figure}

In Fig.~\ref{DMeta1}, we display the correlation between the SI DM–nucleon cross section $\sigma^\text{SI}_{\eta_1}$ and the DM mass $m_{\eta_1}$, corresponding to the parameter space that yields the correct DM relic abundance, $\Om_{\eta_1}h^2\simeq 0.1~\mathrm{pb}/\langle\sigma v_\text{rel}\rangle_{\eta^*_1\eta_1}\simeq 0.12$, as measured by the Planck experiment~\cite{Planck:2018vyg}. The numerical analysis is carried out with the following representative input parameters: $\la_{24}=\la_{25}=\la_3=\la_4=\la'=0.1$, $g_1=g_2=\sqrt2 g_Y$, $v_2=0.52$~GeV, $\La_1=\La_2=10$~TeV, $m_{\phi^0}=2.5$~TeV, and $-\mu_0^2=(0.4\text{--}10)\times 10^4~\mathrm{GeV}^2$, with $Z=54$ and the isotopic composition $(A_i,\rho_i)$ corresponding to natural xenon. Two viable mass regions for $\eta_1$ are identified: $m_{\eta_1}\simeq 592.208$~GeV for $-\mu_0^2\simeq (1.33\text{--}4.26)\times 10^4~\mathrm{GeV}^2$ (left panel), and $m_{\eta_1}\simeq (1.03$--$2.50)$~TeV for $-\mu_0^2\simeq (0.89\text{--}5.28)\times 10^4~\mathrm{GeV}^2$ (right panel). These results are qualitatively similar to those obtained for the $\eta_2$ candidate. However, the projected sensitivities of upcoming direct-detection experiments---XENONnT/LZ~\cite{XENON:2020kmp, LZ:2018qzl}, PandaX-xT~\cite{PANDA-X:2024dlo}, and DARWIN~\cite{DARWIN:2016hyl}---are expected to exclude part, or even all, of the parameter space identified above. To retain viable regions consistent with these future limits, smaller values of the relevant couplings must be considered.

\subsection{Comments on the $\phi^0$ scalar doublet dark matter scenario}
Since $\phi$ is an $SU(2)_L$ doublet with $\mathcal{Y}=1/2-z/3$ and $\mathcal{R}=z/3$, the real and imaginary components of $\phi^0$ couple to the SM $Z$ boson through the interaction $\frac{ig}{2c_W}(S_5\partial^\mu A_5-\partial^\mu S_5A_5)Z_\mu$. If $S_5$ and $A_5$ are degenerate in mass, as assumed in this work, this coupling induces a sizable contribution to the SI direct-detection cross section, implying that $\phi^0$ cannot serve as a viable DM candidate~\cite{Cirelli:2005uq, Barbieri:2006dq}. Nevertheless, we note that $S_5$ and $A_5$ need not be exactly degenerate, as their mixings with other scalar states via the mass matrices $\mathcal{M}^2_{S,A}$ in Eqs.~(\ref{MS}) and (\ref{MA}) can lift the degeneracy. Assuming a mass splitting between $S_5$ and $A_5$ larger than a few hundreds of keV, either $S_5$ or $A_5$ could act as a viable DM candidate, provided it is the lightest among the $P_\mathsf{D}$-odd fields~\cite{Dolle:2009fn, deBoer:2021pon}. 

\section{\label{conclusion}Conclusions}
In this work, we have proposed a novel extension of the SM based on the flipping principle, enlarging the electroweak gauge symmetry with a double right-handed Abelian structure. This framework leads to a chiral charge assignment, where only the right-handed sector carries nontrivial charges, and provides a unified, anomaly-free construction motivated by left-right symmetric theories and grand unified embeddings. 
The model offers a natural explanation for the observed SM fermion mass hierarchy. Third-generation of SM charged fermions acquire tree-level masses via Yukawa interactions with the SM Higgs doublet, while the lighter first- and second-generation masses are radiatively generated at one loop level. This radiative mechanism is mediated by a second scalar doublet, which develops a tiny VEV that is itself generated at one-loop level. For neutrinos, masses arise from a combination of a tree-level type-I and a two-loop radiative seesaw mechanism. A key result is that this structure naturally reproduces the observed neutrino mass hierarchy: the atmospheric mass-squared difference originates at tree level, whereas the solar neutrino mass squared splitting emerges at two-loop level. A residual parity symmetry, which survives spontaneous symmetry breaking, stabilizes the lightest parity-odd scalar, providing a viable DM candidate consistent with the experimental value of DM relic abundance and direct detection constraints. Furthermore, this discrete symmetry enforces the radiative nature of the seesaw mechanisms that generate the masses of the first and second generation of SM charged fermions as well as the solar neutrino mass squared splitting. 
We have examined the phenomenological implications under constraints arising from electroweak precision data, flavor-changing neutral currents, and collider searches. The model predicts new neutral gauge bosons whose signatures are testable at current and future experiments, making the framework both predictive and experimentally accessible.

\section*{Acknowledgement}
AECH is supported by ANID-Chile FONDECYT 1241855, ANID – Millennium Science Initiative Program $ICN2019\_044$, ANID CCTVal CIA250027 and ICTP through the Associates Programme (2026-2031). N.T.Duy was funded by the Vietnam Academy of Science and Technology, Grant No.CBCLCA.03/25-27.

\bibliographystyle{JHEP}

\appendix
\section{\label{poten}Scalar potential}
The scalar potential of our model can be decomposed as $V=V_1+V_2+V_3$, where
\bea V_1 &=& \mu_1^2\Phi_1^\dag\Phi_1 + \mu_2^2\Phi_2^\dag\Phi_2 + \la_1(\Phi_1^\dag\Phi_1)^2 + \la_2(\Phi_2^\dag\Phi_2)^2 + \la_3(\Phi_1^\dag\Phi_1)(\Phi_2^\dag\Phi_2)+ \la_4(\Phi_1^\dag\Phi_2)(\Phi_2^\dag\Phi_1)\crn
&& + \mu_3^2\chi_1^*\chi_1 + \mu_4^2\chi_2^*\chi_2 + \la_5(\chi_1^*\chi_1)^2 + \la_6(\chi_2^*\chi_2)^2  + \la_7(\chi_1^*\chi_1)(\chi_2^*\chi_2)\crn
&& + \chi_1^*\chi_1(\la_8\Phi_1^\dag\Phi_1 + \la_9\Phi_2^\dag\Phi_2) + \chi_2^*\chi_2(\la_{10}\Phi_1^\dag\Phi_1 + \la_{11}\Phi_2^\dag\Phi_2), \\
V_2 &=& \mu_5^2\phi^\dag\phi + \mu_6^2\eta_1^*\eta_1 + \mu_7^2\eta_2^*\eta_2 + \la_{12}(\phi^\dag\phi)^2 + \la_{13}(\eta_1^*\eta_1)^2 + \la_{14}(\eta_2^*\eta_2)^2 + \phi^\dag\phi(\la_{15}\eta_1^*\eta_1 + \la_{16}\eta_2^*\eta_2)\crn
&& + \la_{17}(\eta_1^*\eta_1)(\eta_2^*\eta_2) + \phi^\dag\phi(\la_{18}\Phi_1^\dag\Phi_1 + \la_{19}\Phi_2^\dag\Phi_2 + \la_{20}\chi_1^*\chi_1 + \la_{21}\chi_2^*\chi_2) + \la_{22}(\Phi_1^\dag\phi)(\phi^\dag\Phi_1)\crn 
&& + \la_{23}(\Phi_2^\dag\phi)(\phi^\dag\Phi_2) + \eta_1^*\eta_1(\la_{24}\Phi_1^\dag\Phi_1 + \la_{25}\Phi_2^\dag\Phi_2 + \la_{26}\chi_1^*\chi_1 + \la_{27}\chi_2^*\chi_2)\crn
&& + \eta_2^*\eta_2(\la_{28}\Phi_1^\dag\Phi_1 + \la_{29}\Phi_2^\dag\Phi_2 + \la_{30}\chi_1^*\chi_1 +  \la_{31}\chi_2^*\chi_2) + \mu(\chi_2\eta_1^*\eta_2 + \mathrm{H.c.})\crn
&& + \la[(\Phi_1^\dag\phi)\chi_2^*\eta_2 + \mathrm{H.c.}] + \la'[(\Phi_2^\dag\phi)\chi_2\eta_1 + \mathrm{H.c.}],\\
V_3 &=& \mu_8^2\chi_3^*\chi_3 + \la_{32}(\chi_3^*\chi_3)^2  + \chi_3^*\chi_3(\la_{33}\Phi_1^\dag\Phi_1 + \la_{34}\Phi_2^\dag\Phi_2 + \la_{35}\chi_1^*\chi_1+ \la_{36}\chi_2^*\chi_2) \crn
&& + \chi_3^*\chi_3(\la_{37}\phi^\dag\phi+\la_{38}\eta_1^*\eta_1 + \la_{39}\eta_2^*\eta_2).  \eea
Above, the parameters $\mu$'s have mass dimension, while the $\lambda$'s are dimensionless. Without loss of generality, all of them are taken to be real. To begin with, we impose the conditions $\mu_8^2<0$ and $\la_{32}>0$, and assume a hierarchy $|\mu_8|\gg|\mu_{1,2,\cdots,7}|$ such that the scalar field $\chi_3$ is effectively decoupled from the low-energy dynamics. Under these conditions, $\chi_3$ acquires a large VEV from from the potential term $V_3$, approximately given by $\La_3^2\simeq -\mu_8^2/\la_{32}$, which spontaneously breaks the $U(1)_\mathsf{R}$ symmetry down to a residual parity $\mathsf{P_R}$. 

At energy scales below $\La_3$, integrating out the heavy field $\chi_3$, the effective scalar potential is approximated by $V_\text{eff}\simeq V_1+V_2$. Furthermore, we impose the following conditions:
\be \mu_{1,3,4}^2<0,\hs \mu_{2,5,6,7}^2>0,\hs \la_{1,2,5,6,12,13,14}>0, \hs |\mu_{3,4}|\gg|\mu_{1,2}|, \ee
to ensure that the scalar fields $\chi_{1,2}$ and $\Phi_1$ develop nonzero VEVs, denoted as $\Lambda_{1,2}$ and $v_1$, respectively, with a hierarchy $\Lambda_{1,2} \gg v_1$. In contrast, the scalars $\phi$ and $\eta_{1,2}$ do not acquire VEVs, which helps maintain the residual parity symmetry. These conditions also imply that the VEV of $\Phi_2$ vanishes at tree level, $\langle\Phi_2\rangle=v_2/\sqrt2=0$.

\begin{figure}[h]
\centering
\includegraphics[scale=1.1]{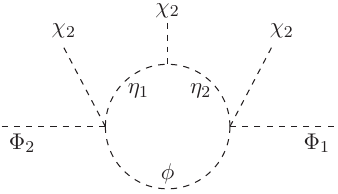}
\caption[]{\label{fig10}One-loop diagram generating the $\mu_0^2\Phi_1^\dag\Phi_2$ term in the scalar potential.}
\end{figure}

However, after the spontaneous breaking of the $U(1)_\mathcal{R}\otimes U(1)_\mathsf{R}$ symmetry, a nonzero VEV of $\Phi_2$ is radiatively induced at the one-loop level. At energy scales below $\La_{1,2}$, the effective potential involving the two scalar doublets $\Phi_1$ and $\Phi_2$ takes the form
\bea V(\Phi_1,\Phi_2) &=& \mu_1^2\Phi_1^\dag\Phi_1 + \mu_2^2\Phi_2^\dag\Phi_2 + \la_1(\Phi_1^\dag\Phi_1)^2 + \la_2(\Phi_2^\dag\Phi_2)^2+ \la_3(\Phi_1^\dag\Phi_1)(\Phi_2^\dag\Phi_2)\crn
&& + \la_4(\Phi_1^\dag\Phi_2)(\Phi_2^\dag\Phi_1) + (\mu_0^2\Phi_1^\dag\Phi_2+\mathrm{H.c.}), \eea
where the effective coupling $\mu_0^2$ arises from the one-loop diagram shown in Fig.~\ref{fig10}, and is given by  
\be \mu_0^2=-\frac{\la\la'\mu\La_2^3}{16\sqrt2\pi^2}\frac{m_{\eta_1}^2m_{\eta_2}^2\ln\frac{m_{\eta_1}}{m_{\eta_2}}+m_{\eta_2}^2m_{\phi}^2\ln\frac{m_{\eta_2}}{m_{\phi}}+m_{\phi}^2m_{\eta_1}^2\ln\frac{m_{\phi}}{m_{\eta_1}}}{(m_{\eta_1}^2-m_{\eta_2}^2)(m_{\eta_2}^2-m_{\phi}^2)(m_{\eta_1}^2-m_{\phi}^2)}.\ee
The masses of the scalar fields entering the loop are given by
\bea m_{\phi}^2 &=& \mu_5^2+\fr 1 2 [(\la_{18}+\la_{22})v_1^2+\la_{20}\La_1^2+\la_{21}\La_2^2],\\
m_{\eta_1}^2 &=& \mu_6^2+\fr 1 2 (\la_{24}v_1^2+\la_{26}\La_1^2+\la_{27}\La_2^2),\\
m_{\eta_2}^2 &=& \mu_7^2+\fr 1 2 (\la_{28}v_1^2+\la_{30}\La_1^2+\la_{31}\La_2^2). \eea
Assuming $\mu_0^2<0$ and $\sqrt{-\mu_0^2}\sim v_1$, the VEV of $\Phi_2$ is approximately given by
\be v_2 \simeq -\frac{2v_1\mu_0^2}{2\mu_2^2+(\la_3+\la_4)v_1^2+\la_9\La_1^2+\la_{11}\La_2^2}, \ee
under the assumption that $v_2\ll v_1$. As a representative example, taking $\la\la'\sim\mathcal{O}(0.1)$, $\mu\sim \mathcal{O}(10)~\mathrm{GeV}$, $\La_2\sim\mathcal{O}(10^4)~\mathrm{GeV}$, and $m_\phi\sim m_{\eta_1}\sim m_{\eta_2}\sim\mathcal{O}(10^3)~\mathrm{GeV}$, we have $-\mu_0^2\sim\mathcal{O}(10^4)~\mathrm{GeV}^2$. Also, for $v_1\sim\mathcal{O}(10^2)~\mathrm{GeV}$, $\mu_2\ll v_1$, $\la_3\sim \la_4\sim\mathcal{O}(0.1)$, $\La_1\sim\La_2\sim\mathcal{O}(10^4)~\mathrm{GeV}$, and $\la_9\sim\la_{11}\sim\mathcal{O}(10^{-2})$, the resulting value $v_2\sim 1$ GeV is sufficient to generate the charm quark mass with an associated Yukawa coupling of order one. Finally, the VEVs $v_{1,2}$ break the electroweak symmetry $SU(2)_L\otimes U(1)_Y$ down to $U(1)_Q$, and simultaneously shift the residual parity $\mathsf{P_R}=(-1)^{2\mathsf{R}}$ to a new conserved parity $\mathsf{P_D}=(-1)^{2\mathsf{D}}$.

\section{\label{matrix}Mass matrix in scalar sector}
To determine the mass matrices in the scalar sector, we expand the scalar fields around their VEVs, namely:
\bea \Phi^0_1 &=& \frac{1}{\sqrt2} (v_1+S_1+iA_1), \hs \Phi^0_2=\frac{1}{\sqrt2} (v_2+S_2+iA_2),\label{eq1}\\
\chi_1 &=& \frac{1}{\sqrt2}(\La_1+S_3+iA_3),\hs \chi_2 = \frac{1}{\sqrt2}(\La_2+S_4+iA_4),\\
\phi^0 &=& \frac{1}{\sqrt2} (S_5+iA_5),\hs \eta_1 = \frac{1}{\sqrt2}(S_6+iA_6),\hs \eta_2 =\frac{1}{\sqrt2}(S_7+iA_7).\label{eq2}  \eea 
Substituting the expanded scalar fields into the effective potential $\mathsf{V}_\text{eff}\simeq V_1+V_2+(\mu_0^2\Phi_1^\dag\Phi_2+\mathrm{H.c.})$, we obtain the conditions for the potential minimum as follows:
\bea 2\mu_0^2v_2+2\mu_1^2v_1+2\la_1v_1^3+(\la_3+\la_4)v_1v_2^2+\la_8\La_1^2v_1+\la_{10}\La_2^2v_1 &=& 0,\\
2\mu_0^2v_1+2\mu_2^2v_2+2\la_2v_2^3+(\la_3+\la_4)v_1^2v_2+\la_9\La_1^2v_2+\la_{11}\La_2^2v_2 &=& 0,\\
2\mu_3^2+2\la_5\La_1^2+\la_7\La_2^2+\la_8v_1^2+\la_9v_2^2 &=& 0,\\
2\mu_4^2+2\la_6\La_2^2+\la_7\La_1^2+\la_{10}v_1^2+\la_{11}v_2^2 &=& 0.\eea
Hence, we determine the mass matrices of scalar bosons. For the CP-even scalar bosons, we obtain two separate mass matrices,
\be M^2_S = \begin{pmatrix} 2\la_1v_1^2-\fr{\mu_0^2v_2}{v_1} & (\la_3+\la_4)v_1v_2+\mu_0^2 & \la_8\La_1v_1 & \la_{10}\La_2v_1 \\ 
(\la_3+\la_4)v_1v_2+\mu_0^2 & 2\la_2v_2^2-\fr{\mu_0^2v_1}{v_2} & \la_9\La_1v_2 & \la_{11}\La_2v_2\\
\la_8\La_1v_1 & \la_9\La_1v_2 & 2\la_5\La_1^2 & \la_7\La_1\La_2\\
\la_{10}\La_2v_1 & \la_{11}\La_2v_2 & \la_7\La_1\La_2 & 2\la_6\La_2^2
 \end{pmatrix}, \ee
 and
\be \mathcal{M}^2_S = \begin{pmatrix} m_\phi^2+\frac{\la_{19}+\la_{23}}{2}v_2^2 & \frac{\la'\La_2v_2}{2} & \frac{\la\La_2v_1}{2} \\ 
\frac{\la'\La_2v_2}{2} & m_{\eta_1}^2+\frac{\la_{25}}{2}v_2^2 & \frac{\mu\La_2}{\sqrt2}\\
\frac{\la\La_2v_1}{2} & \frac{\mu\La_2}{\sqrt2} & m_{\eta_2}^2+\frac{\la_{29}}{2}v_2^2 
 \end{pmatrix},\label{MS} \ee
in the bases $(S_1,S_2,S_3,S_4)$ and $(S_5,S_6,S_7)$, respectively. For the CP-odd scalar bosons, we similarly find two mass matrices,
\be M^2_A = \begin{pmatrix} -\frac{\mu_0^2v_2}{v_1} & \mu_0^2 \\ 
\mu_0^2 & -\frac{\mu_0^2v_1}{v_2}
 \end{pmatrix}, \ee
 and
\be \mathcal{M}^2_A = \begin{pmatrix} m_\phi^2+\frac{\la_{19}+\la_{23}}{2}v_2^2 & -\frac{\la'\La_2v_2}{2} & -\frac{\la\La_2v_1}{2} \\ 
-\frac{\la'\La_2v_2}{2} & m_{\eta_1}^2+\frac{\la_{25}}{2}v_2^2 & \frac{\mu\La_2}{\sqrt2}\\
-\frac{\la\La_2v_1}{2} & \frac{\mu\La_2}{\sqrt2} & m_{\eta_2}^2+\frac{\la_{29}}{2}v_2^2 
 \end{pmatrix}\label{MA}, \ee
in the bases $(A_1,A_2)$ and $(A_5,A_6,A_7)$, respectively. Finally, for the charged scalar bosons, the mass matrix in the basis $(\Phi_1^\pm,\Phi_2^\pm,\phi^\pm)$ takes the form 
\be M^2_C = \begin{pmatrix} -\frac{(\la_4v_1v_2+2\mu_0^2)v_2}{2v_1} & \frac{\la_4v_1v_2}{2}+\mu_0^2 & 0 \\ 
\frac{\la_4v_1v_2}{2}+\mu_0^2 & -\frac{(\la_4v_1v_2+2\mu_0^2)v_1}{2v_2} & 0\\
0 & 0 & m_{\phi}^2+\frac{1}{2}(\la_{19}v_2^2-\la_{22}v_1^2) 
 \end{pmatrix}. \ee

\bibliography{combine}

\end{document}